\documentclass[fleqn,usenatbib]{mnras}

\usepackage{newtxtext,newtxmath}

\usepackage[T1]{fontenc}
\usepackage{ae,aecompl}

\usepackage{graphicx}	
\usepackage{amsmath}	

\title[The mass profile of NGC\,3377]{The mass profile of NGC\,3377 from a Bayesian approach}
\author[J. P. Caso et al.]{Juan P. Caso$^{~1,2}$\thanks{E-mail:
jpcaso@fcaglp.unlp.edu.ar (JPC)}\\
$^{1}$Facultad de Ciencias Astron\'omicas y Geof\'isicas de la Universidad Nacional de La Plata,    
and \\ Instituto de Astrof\'isica de La Plata (CCT La Plata -- CONICET, UNLP), Paseo del Bosque S/N,  
B1900FWA La Plata, Argentina\\   
$^{2}$Consejo Nacional de Investigaciones Cient\'ificas y T\'ecnicas, Godoy Cruz 2290, C1425FQB,  
Ciudad Aut\'onoma de Buenos Aires, Argentina}

\date{Released 2002 Xxxxx XX}

\pagerange{\pageref{firstpage}--\pageref{lastpage}} \pubyear{2002}

\begin{document}

\label{firstpage}

\maketitle

\begin{abstract}
  The mass profile for the moderately bright elliptical NGC\,3377
  is studied through an spherical Jeans
  analysis, combined with a Bayesian approach. The {\it prior}
  distributions are generated from dark matter simulations and
  observational constraints. The observational data set consist of
  Gemini/GMOS long-slit observations aligned with the major and
  minor axes of the galaxy, and are supplemented with data from
  the literature for the diffuse stellar component, globular
  clusters and planetary nebulae. Although the galaxy is assumed
  to alternatively reside in central and satellite haloes, the
  comparison with literature results prefer the latter option.
  Several options of constant anisotropy are considered, as well
  as both NFW and Einasto mass profiles. The analysis points to
  an intermediate mass halo, presenting a virial mass around
  $(3.6\pm0.6)\times 10^{11}\,{\rm M_{\odot}}$. 
\end{abstract}


\section{Introduction} 
\label{intro}

In the current paradigm, it has been generally accepted that cold dark
matter (CDM) is the dominant matter component in the Universe, playing
a main role in the formation of galaxies \citep[e.g.][]{whi91}. The CDM
is assumed to consist of classical, non-relativistic, and collisionless
particles, that only interacts through gravity. From the hierarchical
scenario, the mass growth of haloes is mainly produced by minor mergers,
with a relative contribution of CMD diffuse accretion \citep{wan11}.
Although its relevance in the evolution of the main halo, the fraction
of CDM in subhaloes at $z=0$ is restricted to $\approx 0.1$, from both
dark matter simulations \citep[e.g.][]{spr08,doo14} and semi-analytical
techniques \citep{jia16}.

There is a general consensus from numerical simulation studies that,
once a halo is accreted by a more massive structure, it losses a
considerable fraction of the initial mass \citep{gan10,dra20}, but
the magnitude of this process relative to artificial effects is still
under debate \cite[e.g.][]{vdb18}.
During the first pericentric passage after accretion, subhaloes may
lose $\approx 20-30$ per cent of its mass \citep{rhe17}, increasing
for radial or tightly bound orbits, and for subhaloes with smaller
concentration parameters \citep{ogi19}.
Although complete disruption of subhaloes is extremely rare, successive
passages at the pericenter lead to $\approx 75$ per cent of mass loss
after several Gyr from the infall \citep{nie19}, as well as changes in
the baryonic component of the galaxy \citep{jaf16}, and stripping of
globular clusters, based on both numerical \citep[e.g.][]{cho19,ram20}
and observational studies \citep[e.g.][]{pen08,liu19}. Moreover, the
kinematics for the haloes associated to early-type galaxies (ETGs) point
to diverse properties \citep{pul18}, and some degree of kinematical
complexity in central massive galaxies \citep[][]{ric14,hil18}. 
Then, it is expected to find differences in the behaviour of the mass
relations for central and satellite haloes \citep[e.g.][]{lan16,nie19},
and further analysis of the latter ones is relevant to disentangling
the effect of the environment in the evolution of the mass profile in
satellite galaxies.

From the observational point of view, the measurement of the mass
profiles in early-type galaxies based on kinematical tracers is
challenging in several aspects. The axisymmetric analysis through
integral-field-unit observations of the stellar component is
public for a large sample of galaxies \citep[e.g.][]{cap06,cap13a},
but restricted to galactocentric distances in the range of $\approx 1-2$
effective radii of the stellar component. With the exception of
some peculiar cases \citep[e.g.][]{lan15}, for larger distances
the analysis is usually run through halo tracers, i.e. globular
clusters (GCs) and planetary nebula (PNe). These studies assume
spherical symmetry for the mass distribution, and involve hundreds
of tracers \citep[][]{nap11,sch12,ric15}, restricting the analysis
to massive galaxies, that typically present large populations of
halo tracers \citep[e.g.][]{cor13,har17}. But satellite galaxies
usually present rather poor halo populations, and the size of the
samples is also restricted by observational constraints on the
feasible signal-to-noise to be achieved in the observations. On
the basis of different assumptions, a variety of mass estimators
are used in these cases, with different degrees of uncertainties
\citep{cas14,ala16,ko20}.

The Leo I is a nearby group of galaxies, conformed by a main body,
dominated by M\,96, and the Leo triplet, which is about six degrees
to the East. The M\,96 subgroup contains seven bright galaxies,
including NGC\,3377, and a dwarf population \citep{mul18}. An initial
photometric catalogue was built by \citet{fer90}, and subsequently
enlarged by other studies \citep[e.g.][]{kar04a,coh18}, including
analysis of the velocity space to identify members and background
objects \citep[e.g.][]{tre02,sti09}. \citet{fli03} report a gap
in the luminosity function of the group for intermediate brightness
galaxies.
One of its main features is the `Leo Ring' \citep{sch83}, an
intergalactic HI complex
that surrounds the galaxies NGC\,3384 and M\,105, and presents a
stream that also connects it with M\,96. Although, \citet{wat14}
indicate that interaction signatures in the central part of the
M\,96 subgroup are subtle for this type of environment, without a
significant contribution of diffuse intragroup starlight. This
points that encounters are relatively mild and infrequent. It
seems a dynamically relaxed system, with bright galaxies more
concentrated towards the centre of the group than the dwarf
population \citep{kar04b}. 

The galaxy NGC\,3377 is a moderately bright elliptical,
with elongated morphology \citep[E5-6][]{dev91} and kinematics
up to the effective radius corresponding to a fast rotator
\citep{ems11}. \citet{coc09} concluded that the galaxy presents
disc-like kinematics up to $\approx 70$\,arcsec, with high values
of velocity rotation and depressed velocity dispersion for both
the stellar component and PNe populations along the semi-major
axis, besides disky isophotes. For outer radii, the kinematical
behaviour changes, pointing to a fading of the disc component.
The photometric properties of the globular cluster system (GCS)
were studied by \citet{cho12} and \citet{cas19}, pointing to an
old population, with bimodal colour distribution and an extension
that agrees with the typical values found in galaxies with similar
luminosity. The kinematics of the GCS were studied by \citet{pot13},
who reported significant rotation in the red GCs, but no evidence
for the blue ones. In the following we will assumed for NGC\,3377
a distance of $10.6\pm0.4$\,Mpc \citep{tul13}.

\smallskip
The aim of this study is to constraint the mass profile of NGC\,3377,
providing an alternative procedure to mass estimators which could
be useful in the analysis of intermediate-mass ellipticals.
The paper is organised as follows. The observational and numerical
dataset are described in Section\,2. The procedure is described
in Section\,3 and the results are presented in
Section\,4. Section\,5 is devoted to compare the results with
mass estimations and several scaling relations from the literature.
Finally, in Section\,6 is presented a brief summary.

\section{Observational and numerical data}
\label{sec.obs}

\begin{figure}
  \centering
\includegraphics[width=70mm]{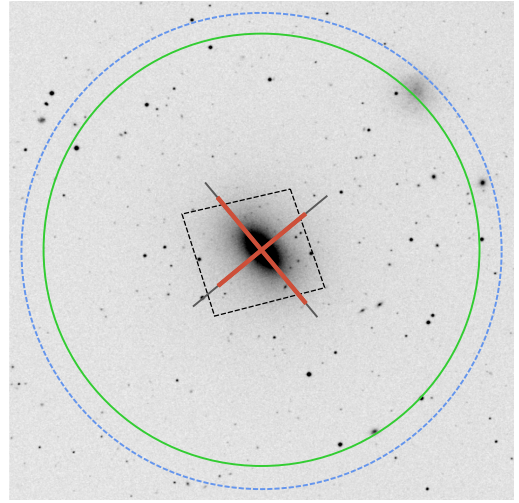}    
\caption{Image in the $R$ filter centred on NGC\,3377, obtained
  through the ESO Digitized Sky Survey. The FOV is
  $16 \times 16$\,arcmin$^2$, and it follows the usual orientation
  (North is up, East to the left). The black dashed polygon represents
  the HST/ACS FOV. The thin grey lines correspond to the position
  and length of the long-slits observed through Gemini/GMOS-N, while
  the thick red ones highlight the radial range from which are
  measured the kinematical data. The green solid circle represents
  for comparison purposes the region containing the PNe from
  \citet{coc09}, and the blue dashed circle corresponds to GCs from
  \citet{pot13}.}    
\label{field}    
\end{figure}

\subsection{Surface brightness profile}
\label{photgal}
For the surface brightness profile of NGC\,3377, a S\'ersic
law \citep{ser68} is assumed, with the parameters fitted by
\citet{kra13} to the $r$ filter, and converted to equivalent
radius ($R_{\rm eq}$) with the axes ratio, $q_{\rm ab}=0.52$. The
absorption in
the $r$ filter is obtained from NED\footnote{This research 
has made use of the NASA/IPAC Extragalactic Database (NED) 
which is operated by the Jet Propulsion Laboratory, California
Institute of Technology, under contract with the National
Aeronautics and Space Administration.}, and corresponds to
$A_r=0.077$\,mag, following the calibration by \citet{sch11}.
In order to convert the units from mag\,arcsec$^{-2}$ to
${\rm L_{\odot} pc^{-2}}$, it is used $M_{r,\odot}=4.65$\,mag
\citep{wil18} for the absolute magnitude of the sun, and
the distance already stated in Section\,\ref{intro}. Hence,
the expression results

\begin{equation}
  I_r(R_{\rm eq}) = I_0\,{\rm exp} \left[-\left(\frac{R_{\rm eq}}{R_0}\right)^{\frac{1}{n_r}}\right]
\label{eq.sersic}
\end{equation}

\noindent using the approximation of $b_{\rm n}$ from \citet{cio91}, with
$n_r=5\pm0.5$, $I_0= (1.184\pm0.2)\times10^6\,{\rm L_{\odot} pc^{-2}}$ and $R_0= 0.023\pm0.01$\,pc.
Further analysis in this paper requires to deproject the S\'ersic law.
Following the approximation from \citet{pru97} and \citet{lim99}, see also
\citet{sal12}, it results

\begin{equation}
    \begin{split}
      j(x) & \simeq j_1\left(\frac{x}{R_0}\right)^{-p} {\rm exp}\left[-\left(\frac{x}{R_0}\right)^{\frac{1}{n_r}}\right], \\
        j_1  & = \frac{\Gamma (2n_r)}{\Gamma\left[(3-p)n_r\right]}\frac{I_0}{R_0}, \\
        p & \simeq 1 - \frac{1}{n_r} + \frac{0.05463}{n_r^2} \\
    \end{split}
\end{equation}

The stellar mass enclosed at a radius $x$ emerges from the integration of
the previous expression and the mass-to-light ratio, which is assumed
$M/L_r=3.55\pm0.2$ by \citet{cap13b}. For the cumulative luminosity, we follow
\citet{lim99},

\begin{equation}
    \begin{split}
      L_r(x) & = L_{\rm tot}\frac{\gamma\left[(3-p)n_r,(x/R_0)^{\frac{1}{n_r}}\right]}{\Gamma\left[(3-p)n_r\right]}, \\
      L_{\rm tot} & = 2\pi n_r \Gamma(2n_r)I_0 R_0^2 \\
    \end{split}
\end{equation}

\begin{figure}  
\includegraphics[width=85mm]{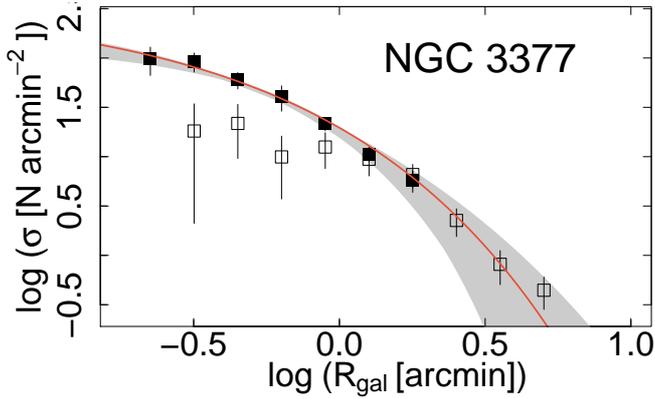}    
\caption{Projected radial distribution for the GCS of NGC\,3377,
  corrected by completeness and contamination. The grey region
  represents the uncertainties in the S\'ersic profile, and the red
  solid curve corresponds to the resulting fitting. The open
  symbols shows the projected density of the spectroscopically
  confirmed sample from \citet{pot13}.}    
\label{GCrad}    
\end{figure}

\subsection{Globular clusters candidates}
\label{photGCs}
The photometric data set corresponds to a single field, centred on
NGC\,3377 (indicated with dashed lines in Fig.\,\ref{field}), and
observed through the HST/ACS Wide Field Camera,
during January 2006 (programme ID 10554). Both raw and processed
images are available at the Mikulski Archive for Space Telescopes
(MAST)\footnote{Based on observations made with the NASA/ESA Hubble
Space Telescope, obtained from the data archive at the Space
Telescope Science Institute. STScI is operated by the Association
of Universities for Research in Astronomy, Inc. under NASA
contract NAS 5-26555.}. These observations correspond to the filters
$F475W$ and $F850LP$, commonly used in extragalactic GC studies,
and the total exposure times are 1380\,s and 3005\,s, respectively.
Its field-of-view (FOV) is $202 \times 202$\,arcsec$^2$.

In the following, the procedure for detecting and measuring the GC
candidates is summarised. First, the diffuse galaxy light is
subtracted in order to improve the detection of sources close to
the galaxy centre in both filters. This is achieved by means of
several tasks from {\sc iraf} (version V2.16). First, elliptical 
isophotes are fitted to the galaxy in each filter using the task
ELLIPSE. Bright stars and background galaxies on the field are masked
to achieve a stable solution. The parameters ellipticity and position 
angle are simultaneously fitted for the inner $\approx 45$\,arcsec. 
For larger galactocentric distances, the position angle is fixed to 
avoid fluctuations due to the low surface brightness level and the 
edges of the FOV. Then, this output is used to modelled the diffuse 
galaxy light with the task BMODEL, and finally subtracted to the 
original image. With the exception of some underlying features from
the disk component in the inner 15\,arcsec, the diffuse galaxy light
is removed from the images. This procedure has proven to favour the 
detection of GC candidates in the inner regions of early-type galaxies 
in previous papers including ACS data \citep[e.g.][]{cas13,cas19,deb22}.
Afterwards, SE{\sc xtractor} \citep{ber96} is run in both filters to
build up a preliminary catalogue of sources, assuming as a positive
detection every group of at least three connected pixels with counts
above the local sky level plus $3\sigma$. Although GCs at the assumed
distance might be marginally resolved, from a typical effective radius
of $\approx 3$\,pc \citep[e.g.][]{jor05,bro11,cas14}, and assuming low
eccentricities \citep[e.g.][]{har09}, their full width at half-maximum
(FWHM) should not exceed a few pixels. In order to reject extended
sources, only those sources presenting FWHM smaller than 5\,px and
elongation smaller than 2 in both filters are selected, criteria also
used in other studies with similar instrumental configuration and
distances to the galaxies \citep[e.g.][]{jor04,jor07}. Aperture
photometry is performed in both filters with an aperture radius of
5\,px, and mean aperture corrections are calculated from the analysis
of structural parameters of marginally resolved sources with ISHAPE
\citep[][see \citealt{cas19} for further details]{lar99}.
The photometry is calibrated to $g$ and $z$ filters, applying the
zero-points from \citet{sir05}, $ZP_{F475}=26.068$ and $ZP_{F850}=24.862$\,mag.
Then, the magnitudes are corrected by extinction, assuming the values
from NED, $A_g=0.11$\,mag and $A_z=0.04$\,mag \citep{sch11}. 

Finally, GC candidates are chosen from those sources with colours
fulfilling $0.6 < (g-z)_0 < 1.7$\,mag, similar colour range than those
assumed in previous GC studies using these bands \citep[e.g.][]{jor05}.
Completeness functions for point-like sources are calculated from
artificial stars in different galactocentric distances, to account
for variations due to Poisson noise from the galaxy surface brigthness.
Then, the faint limit for the photometric analysis is set at $z=24$\,mag.
This assures a high completeness level, and spans $\approx 95\%$ of the
GCs, based on the GCs luminosity function from \citet{cho12}.
We refer to \citet{cas19} for further information about the calculus
of the completeness correction for GC candidates in similar images.

\begin{figure*}    
\includegraphics[width=160mm]{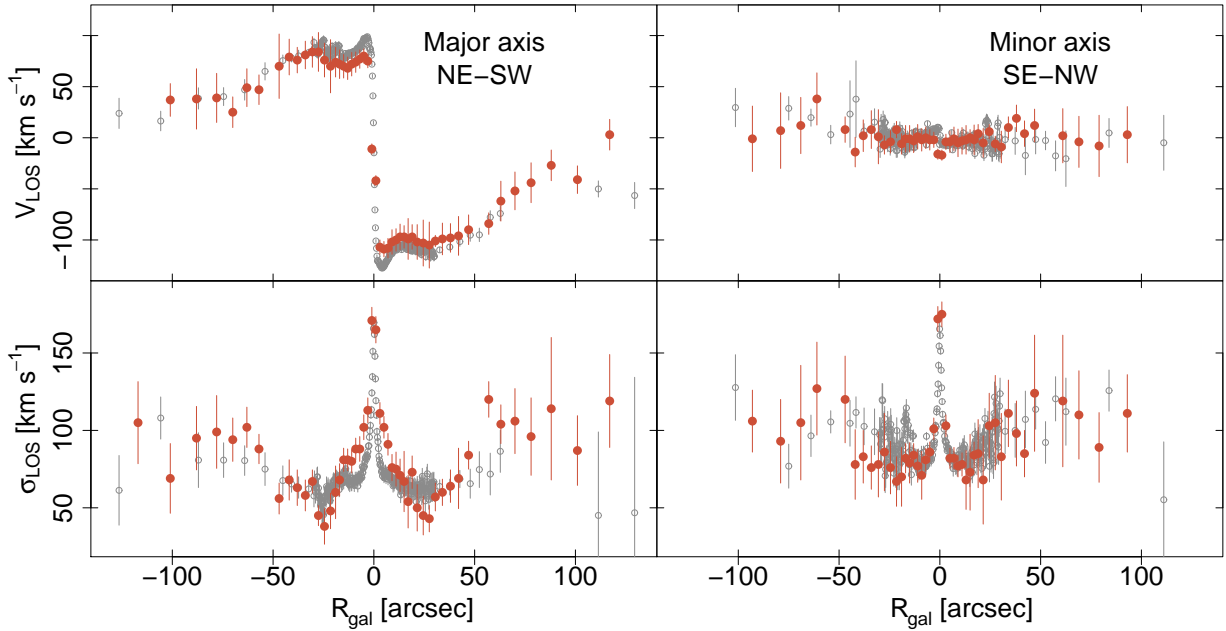}    
\caption{Long-slit stellar kinematics of NGC\,3377, extracted along the
photometric major (left-hand panels) and minor (right-hand panels) axes.
The upper panels correspond to the velocity in the line-of-sight
($V_{\rm LOS}$), corrected by the systemic velocity of NGC\,3377. The lower
panels represent the velocity dispersions ($\sigma_{\rm LOS}$). For
all panels, red filled circles indicate data presented in this paper, and
grey open circles are from \citet{coc09}.}
\label{kinem}    
\end{figure*}

Figure\,\ref{GCrad} shows the radial distribution for the GCS of
NGC\,3377 as a function of the projected galactocentric distance
($R_{\rm gal}$), corrected by completeness and subtracted by
contamination. The completeness corrections are obtained from
completeness functions at four galactocentric ranges, to
consider the changes on GCs detection as a consequence of the
increasing noise close to the galaxy centre. The contamination is
assumed $n_{\rm b}= 1$\,arcmin$^{-2}$, following \citet{cho12} for
the same observations and photometric depth. The radial binning
is constant on logarithmic scale, $log_{10} \Delta R{\rm _{gal}\,[arcsec]}=7$.
A S\'ersic profile (see Equation\,\ref{eq.sersic}) is fitted for different
bin breaks, slightly shifted to consider discreteness issues. The grey
region in the figure represents the changes in the S\'ersic profiles,
while the red solid curve corresponds to the weighted mean of the
parameters fitted in each individual run. These result $n_{\rm GCs}=3.1$,
$I_{\rm 0,GCs}=3900$\,arcmin$^{-2}$ and $R_{\rm 0,GCs}=0.0052$\,arcmin. The
open symbols double the projected density for the spectroscopically
confirmed GCs brighter than $g'=23$\,mag, from \citet{pot13}, for
comparison purposes. This latter magnitude limit corresponds to the
turn-over magnitude of the GCS, according to \citep{cho12}. For the
outer region both profiles agree.
The extension of the GC system (GCS) is assumed $7.5$\,arcmin, that
corresponds to the extension of the spectroscopic survey by \citet{pot13}
and does not differ significantly from the extension of $6.2$\,arcmin
calculated by \citet{cas19}. The numerical integration of this profile
up to $7.5$\,arcmin results $325\pm30$ GCs, which doubles previous
results from ACS studies \citep{cho12}.

\subsection{Gemini/GMOS spectroscopy}
\label{obs.gmos}
Long-slit spectra of NGC\,3377 has been observed with the Gemini
Multi-Object Spectrograph (GMOS) mounted on Gemini North, during
January/February in 2021 (GN-2020B-Q-401, PI Juan Pablo Caso).
The aim of these observations was to complement data from the literature 
at large galactocentric distances \citep[e.g.][]{coc09}. Hence they
were carried on relaxed observing conditions, with typical seeing of 
$\approx 2$\,arcsec.
The observations comprise two orientations, following the position
angles for the photometric major and minor axes of the stellar component
(PA$\approx 40\deg$ in the NE-SW direction, and PA$\approx 130\deg$
in the SE-NW direction, respectively, and represented with grey
solid lines in Fig.\,\ref{field}). The observations are split in
16 exposures of 900\,s each. The chosen grism is B600$\_$G5307,
with a slit width of 1\,arcsec, $2\times 2$ binning and central
wavelengths in the range $5400-5550$\,\r{A}, to account for CCD gaps.
The resulting spectral resolution is $\approx 6$\,\r{A}.

Both flatfields and CuAr lamps has been observed with each science
image, to account for flexion effects. The standard reduction process is
performed with the Gemini-GMOS routines in {\sc iraf} (version V2.16).
The two-dimensional spectra are spatially rebinned to achieve a
$S/N \gtrsim 20$ at 5000\,\r{A}. Considering the typical seeing
during the observations, the minimum spatial bin is set at 2\,arcsec.
The individual spectra are extracted in the wavelength range $4800-5800$\,\r{A}, 
that contains several absorption features typical of the spectra in elliptical 
galaxies (H$\beta$, Mgb, and Fe lines). The mean velocity and velocity 
dispersion in the line-of-sight are measured by means of the penalised Pixel 
Fitting code \citep[p{\sc pxf},][]{cap04,cap17}. A subset of the MILES single
stellar population models \citep{vaz10} are selected as templates for
p{\sc pxf} fitting, considering old populations (8 and 10\,Gyrs) and
a wide range of metallicities (${\rm [Z/H]=}$ -1.31, -0.71, -0.4 and 0.0).
The results for both slit orientations are listed in Table\,\ref{tab.kinem}
at the Appendix. For the long-slit corresponding to the major axis
of NGC\,3377, kinematical measurements are obtained up to $\approx2$\,arcmin,
and up to $\approx1.6$\,arcmin for the minor one. Both radial ranges are 
represented by red thick lines in Figure\,\ref{field}.

Figure\,\ref{kinem} shows the velocity ($V_{\rm LOS}$) and velocity
dispersion in the line-of-sight ($\sigma_{\rm LOS}$) as a function of
$R_{\rm gal}$, expressed in arcsec, for the major (left-hand panels)
and minor axes (right-hand panels). Filled red circles correspond
to this paper. For comparison purposes, grey open symbols refer to
the previous kinematical analysis performed by \citet{coc09}. The
measurements from both datasets are in acceptable agreement, taking
into account uncertainties and the differences in the spatial binwidth 
for the inner 30\,arcsec. From a systemic
$V_{\rm LOS}$ for NGC\,3377 of $690\pm5$\,km\,s$^{-1}$ \citep{cap11},
the rotation velocity along the major axis achieves
$\approx 100$\,km\,s$^{-1}$ up to 40\,arcsec from the galaxy centre,
decreasing outwards. Besides, $\sigma_{\rm LOS}$ rapidly decreases from
$\approx 180$\,km\,s$^{-1}$ to a minimum of $\approx 60$\,km\,s$^{-1}$ at
40\,arcsec, to increase up to $\approx 100$\,km\,s$^{-1}$ in the outskirts.
On the contrary, along the minor axis $V_{\rm LOS}$ experiences a flatten
distribution, and the profile of $\sigma_{\rm LOS}$ is soften. From these,
\citet{coc09} conclude that the galaxy presents a disc-like kinematics in
the inner $\approx 60$\,arcsec, with a larger contribution of the bulge,
dynamically hotter, along the minor axis.

\begin{figure}    
\includegraphics[width=85mm]{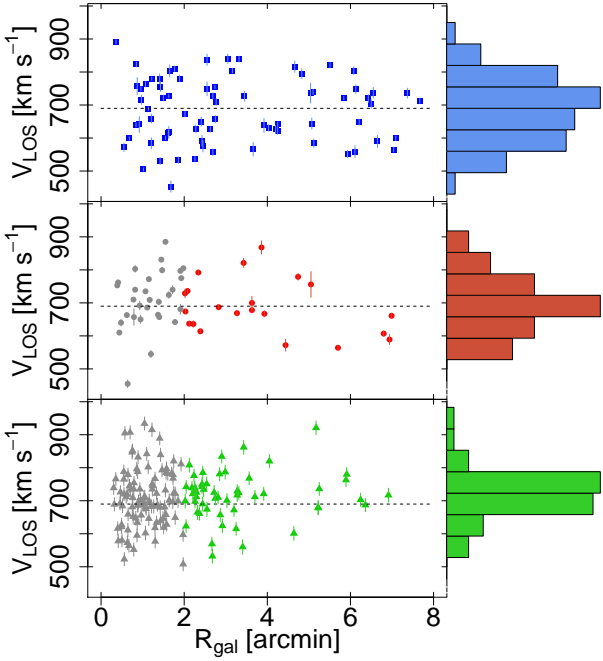}    
\caption{Velocity in the line-of-sight ($V_{\rm LOS}$) as a function of
the galactocentric distance ($R_{\rm gal}$) for blue (upper panel) and
red GCs (middle panel), and PNe (lower panel). The horizontal dashed
line corresponds to the systemic $V_{\rm LOS}$ of NGC\,3377. The objects
in grey colour are not used in the present work due to signs of
rotation.}    
\label{dvel}    
\end{figure}

\subsection{Other kinematical sources}
\label{sec.kin}

From the SLUGGS survey, \citet{pot13} present 126 spectroscopically
confirmed GCs for NGC\,3377. The sample spans up to $\approx7.5$\,arcmin
from the galaxy centre (represented by the blue dashed circle in
Fig.\,\ref{field}), and their uncertainties are typically
${\rm \approx 5-15\,km\,s^{-1}}$. The authors indicate that blue GCs
do not present evidence of rotation, but red GCs are consistent with
a rotation velocity $\approx 50-60$\,km\,s$^{-1}$. For $R_{\rm gal}$
larger than 2\,arcmin, the evidence for rotation becomes less accurate
due to the size of the sample and the bias estimation presented for the
method in their appendix. In this paper, the colour limit for blue and
red GCs is assumed at $(g-i)=0.9$. The upper panel from Figure\,\ref{dvel}
shows the $V_{\rm LOS}$ for blue GCs as a function of $R_{\rm gal}$, while the
middle panel is analogue for red GC. For the latter ones, only objects
presenting $R_{\rm gal}> 2$\,arcmin are used in this study. Outlier objects
are rejected iteratively, considering a $3\sigma$ criteria. Then,
the final sample of GCs contains 93 objects, and its mean $V_{\rm LOS}$
results $685\pm9$\,km\,s$^{-1}$, which matches with the systemic
$V_{\rm LOS}$ of NGC\,3377. Following the adjusted Fisher-Pearson estimator
for the kurtosis excess \citep{joa98}, it results $-0.8\pm0.4$. The
uncertainty is approximated under the assumption of normality. 

The planetary nebulae (PNe) in NGC\,3377 have been studied by \citet{coc09},
resulting in a spectroscopically confirmed sample of 154 objects up to
$\approx 6$\,arcmin (represented by the green solid circle in
Fig.\,\ref{field}). They also indicate evidence of rotation in the PNe
population, with rotation velocity and kinematic position angle in
agreement with the stellar population of the galaxy. For galactocentric
distances larger than 2\,arcmin, the mean velocity rotation seems to
decrease to values consistent with zero in both the major and minor
axes \citep[see Figure\,7 from][]{coc09}. Then, those PNe presenting
$R_{\rm gal}> 2$\,arcmin are selected for the present study. After the
rejection of outliers, the sample consists of 53 objects with a mean
$V_{\rm LOS}$ of $711\pm14$\,km\,s$^{-1}$. 
The kurtosis excess and its approximated uncertainty
result $0.8\pm0.6$. The mean $V_{\rm LOS}$ of the PNe differs from
the systemic velocity of NGC\,3377 in $\Delta V_{\rm LOS} = 21\pm15$
km\,s$^{-1}$, and in $\Delta V_{\rm LOS} = 26\pm17$\,km\,s$^{-1}$ from 
the mean $V_{\rm LOS}$ of the CGs. A revision of both samples do not 
lead to any particular feature to explain this differences. In the range 
$2 < R_{\rm gal} < 2.5$\,arcmin, 16 of 20 PNe present $V_{\rm LOS}$ 
larger than $690$\,km\,s$^{-1}$, but the distribution of their position 
angles does not seem to follow the rotation of the inner disk. A thousand
Monte-Carlo simulations are run, for normal samples with equal mean, 
dispersion in the range of $75-85$\,km\,s$^{-1}$, and sizes corresponding 
to those of GCs and PNe samples used in the present work. 
Observational uncertainties are added to the objects, following the 
observed ones. It results in $\approx 12$ per cent of the cases presenting 
$\Delta V_{\rm LOS} > 25$\,km\,s$^{-1}$. Then, it cannot be concluded 
that the result implies some systematic difference between both samples.

\subsection{Dark matter haloes from a numerical simulation}
\label{mdpl2}
It is also used the cosmological dark matter simulation MDPL2, part
of the Multidark project
\citep[][and publicly available through the official database of
the project\footnote{\url{https://www.cosmosim.org/}}]{kly16}.
This simulation spans a periodic cubic volume of $1\,h^{-1}\,{\rm Gpc}$
of size length. It contains $3840^3$ particles with mass of $1.51
\times10^9\,h^{-1}\,{\rm M_\odot}$ and it considers the cosmological
parameters of \citet{pla13}. The dataset consists of the catalogue
of dark matter haloes detected with the {\sc rockstar} halo finder
\citep{beh13} in the snapshot that corresponds to the local Universe
($z=0$). Each halo is assumed to be the host of a unique galaxy, with
the main halo hosting the central galaxy of each system, while the
satellite haloes host the satellite galaxies.

In addition to the properties described in the catalogue, a luminosity
in the $r$ band is assigned to each halo in a non parametric way, by
means of a halo occupation distribution method \citep[HOD][]{val06,con06}.
Under the assumption of a monotonic relation between the galaxy
luminosities and the halo virial masses, the number density of haloes
in a mass range match the number density of galaxies in the analogue
luminosity range. While the first quantity is obtained from the
simulation, the latter one follows the galaxy luminosity function (LF),
described through a Schechter function \citep{sch76}. The parameters of
the LF correspond to the fits available in \citet{lan16}, that
discriminate between central and satellite galaxies.
The $M_r$ magnitudes are assigned in decreasing order of luminosity,
assuming a constant step of $0.01$\,mag. Observational and numerical
studies have pointed in the literature to an intrinsic scatter in the
stellar-to-halo mass relation \citep[SHMR, e.g.][]{erf19,leg19,mos10},
that is usually represented by a log-normal distribution. Although the
dependence with halo mass has been pointed by some authors
\citep[e.g.][]{mat17}, in this work is assumed a constant scatter of
$0.20$ following \citet{gir20}. Hence, the final $M_r$ magnitudes for
the haloes account for this scatter through the latter expression.
We are aware that inclusion of baryonic matter leads to some tension
between cosmological simulations and observations from the Local Group
in the low mass regime \citep[][and references there in]{web20}, but
the expected mass for NGC\,3377 is beyond this range.

\begin{figure}    
\includegraphics[width=85mm]{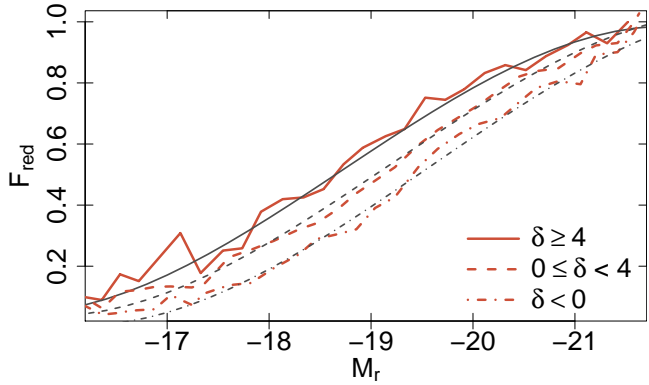}    
\caption{Fraction of red galaxies from \citet{mcn14} as a function of $M_r$
  (thick red lines) and third-order polynomials fitted to these relations
  (thin grey curves), for three different ranges of environmental density.
  The parameter $\delta$ corresponds to the normalised density of neighbours
  brighter than $M_{r,{\rm lim}}= -21$\,mag (see text).}
\label{morf}    
\end{figure}

Finally, the morphological type of the galaxy hosted in each halo is
randomly selected between early and late type, on the bases of the red
fraction of galaxies in different environments from \citet{mcn14}. To
do so, a numerical density is calculated for each halo, $n_{\rm neig}$,
defined as the number of neighbours closer than $8\,h$\,Mpc that present
luminosities brighter than $M_{r,{\rm lim}}= -20.1 + 5\,{\rm log_{10}}(h)$,
i.e. $M_{r,{\rm lim}}= -21$\,mag, due to completeness effects in the sample
of \citet{mcn14}. Third-order polynomials are fitted to the fraction of
red galaxies as a function of $M_r$
\citep[see Figure\,11 from][and Figure\,\ref{morf} in this paper]{mcn14},

\begin{equation}
  \begin{split}
    F_{\rm red}(x) & = 44.5 + 7.48x + 0.41x^2 + 0.01x^3,\,{\rm for }\,\delta < 0 \\
      &  = 44.1 + 7.26x + 0.39x^2 + 0.01x^3\,{\rm for }\,0 < \delta < 4 \\
      &  = 45.9 + 7.41x + 0.39x^2 + 0.01x^3\,{\rm for }\,\delta > 4 \\
  \end{split}
\end{equation}

\noindent with $x$ corresponding to the $M_r$ assigned to each halo, and
$\delta$ to the environmental density, $n_{\rm neig},$ normalised by its
mean value, $\delta = (n_{\rm neig}- \overline{ n_{\rm neig}})/\overline{ n_{\rm neig}}$

These polynomials and the previously derived numerical densities are used
to separate the haloes hosting red and blue galaxies, as a proxy of early
and late-type ones.

\section{Dynamical modeling}
\label{sec.model}

\subsection{Spherical Jeans analysis}
\label{sec.jeans}
Previous dynamical analysis up to the effective radius \citep{cap13a,cap13b}
assumes axisymmetric models for NGC\,3377, which is classified as a fast rotator at 
that radial range. In this case, we adopt a different scenario for the outskirts of the
galaxy, based on both spectroscopical and photometrical data. In Section\,\ref{obs.gmos}
it has been already noticed that the behaviour of $V_{\rm LOS}$ and $\sigma_{\rm LOS}$ 
along the major axis outwards 40\,arcsec points to a decreasing contribution of the
disc component. \citet{arn14} analyse the two-dimensional photometry of NGC\,3377,
resulting in boxy isophotes and decreasing ellipticity outwards $\approx 30$\,arcsec. 
The authors decompose the galaxy surface brightness profile into three components, 
pointing to an increasing contribution of the bulge for galactocentric distances 
larger than $\approx 30$\,arcsec. Assuming constant velocity rotations for each
component, it leads that the kinematics of the slow-rotating bulge dominates over 
the inner disk outwards $60-70$\,arcsec. Then, we assume that outwards 60\,arcsec of
galactocentric distance, NGC\,3377 behaves as a non-rotating systems with spherical 
symmetry, and the results from \citep{cap13a,cap13b} for the inner region are
considered as constraints for the {\it prior} distribution (see Section\,\ref{sec.prior}).
On the basis of the previous assumptions, the second-order velocity moments are 
related by the Jeans equation, \citep[e.g.][]{lok02,bin08}, resulting:

\begin{equation}
  \frac{d\left[j(r){\sigma_r}^2(r)\right]}{dr} + \frac{2\beta}{r}j(r){\sigma^2_r}(r) = -j(r)\frac{GM(r)}{r^2}
\end{equation}

\noindent with $j(r)$ as the three-dimensional density of the tracer
population, $\beta$ is the anisotropy parameter, that accounts for
possible departures from pure isotropy ($\beta>0$ for radial orbits,
and $\beta<0$ for tangential ones), ${\sigma_r}^2(r)$ is the radial
component of the velocity dispersion at the spatial galactocentric
distance $r$, and $M(r)$ is the enclosed total mass.
In the case of the fourth-order moments, additional assumptions are
needed to constraint the solutions. Adopting a distribution function of
the form $f(E,L)=f_0(E)L^{-2\beta}$, with constant anisotropy parameter,
the ratio of the fourth-order moments are related with $\beta$, and the
Jeans equations are reduced to a single one of the form (\citealt{lok02},
see also \citealt{ric13}):

\begin{equation}
  \frac{d\left[j(r)\overline{{V_r}^4}(r)\right]}{dr} + \frac{2\beta}{r}j(r)\overline{{V^4_r}}(r) = -3j(r)\sigma^2_r\frac{GM(r)}{r^2}
\end{equation}

\noindent with $\overline{{V_r}^4}(r)$ the radial component of the
fourth-order moment of the velocity. Then, the solution to previous
equations, assuming constant $\beta$, reads: 

\begin{equation}
  \label{sigmar}
  j(r){\sigma_r}^2(r) = \int_r^{\infty}j(s)\frac{G\,M(s)}{s^2}\left(\frac{s}{r}\right)^{2\beta}ds
\end{equation}

\begin{equation}
  \label{kurr}
  j(r)\overline{{V_r}^4}(r) = 3\int_r^{\infty}j(s){\sigma_r}^2(s)\frac{G\,M(s)}{s^2}\left(\frac{s}{r}\right)^{2\beta}ds
\end{equation}

From the observational perspective, the projections of the velocity
moments represent more useful parameters, leading to direct comparison
with measurable kinematics. The integration along the line-of-sight
results:

\begin{equation}
  \label{sigmalos}
  \sigma_{\rm LOS}^2(R) = \frac{2}{n(R)}\int_R^{\infty}\left(1-\beta\frac{R^2}{r^2}\right) \frac{j\sigma_r^2r dr}{\sqrt{r^2-R^2}}
\end{equation}

\begin{equation}
  \label{kurlos}
  \overline{V_{\rm LOS}^4}(R) = \frac{2}{n(R)}\int_R^{\infty}\left[1-2\beta\frac{R^2}{r^2}+\frac{\beta(1+\beta)}{2}\frac{R^4}{r^4}\right]\frac{j\overline{V_r^4}r dr}{\sqrt{r^2-R^2}}
\end{equation}

\noindent with $\sigma_{\rm LOS}(R)$ and $\overline{V_{\rm LOS}^4}(R)$ the
second and fourth-order moments in the line-of-sight, respectively, $n(R)$
the projected density of the tracer population, and $R$ the projected
radius. To simplify the previous expressions, it is avoided to make
explicit the dependence on the three dimensional radius, $r$, of the
stellar density and the moments of the velocity. Introducing
Equation\,\ref{sigmar} into \ref{sigmalos} and inverting the order of
integration, it yields the integral solution

\begin{equation}
  \sigma_{\rm LOS}^2(R) = \frac{2}{n(R)}\int_R^{\infty}K(R,r)j(r)M(r)\frac{dr}{r}
\end{equation}

\noindent where the kernels $K(R,r)$ vary for different values of
constant anisotropy, as indicated at the Appendix from \citet{mam05}.
From Equations\,\ref{sigmar} and \ref{kurr}, Equation\,\ref{kurlos}
results in a triple integral. Inverting the order of integration
and rearranging the factors, it is obtained a solution of the form:

\begin{equation}
\begin{split}
  \overline{V_{\rm LOS}^4}(R) & =  \frac{2}{I(R)} \int_R^\infty du\,G\,M(u)\,j(u)\,r^{2\beta-2} \\
  &  \int_R^u\,ds\frac{G\,M(s)}{s}s^{-2\beta}\,K^4({R,s})\\
\end{split}
\end{equation}

\noindent with analogue kernels $K^4(R,s)$, presented at the Appendix\,A
for constant anisotropy, ranging $-1/2 \leq \beta \leq 1/2$.

\subsection{Bayesian treatment}
\label{bayes}
This work applies a Bayesian approach to estimate the mass profile of
the galaxy. The components of the observational dataset $D$ have been
already described in Section\,\ref{sec.obs}. It can be split in the
density profiles of the tracer populations,
$D_{\rm rad} =\{n_{\rm Gal},n_{\rm GCs}\}$, and the kinematical data,
$D_{\rm kin}\,=\,\{{\rm Gal,GCs,PNe}\}$.

Several popular models used to describe dark matter haloes with spherical
symmetry can be described by two parameters \citep[e.g.][]{bur95,nav97},
typically a scale radius and a density or concentration parameter, for
instance $r',\rho'$. Hence, the Bayesian analysis points to finding the
probability $p\left(r',\rho'|D,I_{\rm prior}\right)$, with $I_{\rm prior}$
representing the information not included in the dataset that will mould
the {\it prior} distribution. From the Bayes theorem,

\begin{equation}
\label{eq.bayes}
p\left(r',\rho'|D,I_{\rm prior}\right) =
    \frac{p\left(r',\rho'|I_{\rm prior}\right)\,p\left(D|
      r',\rho',I_{\rm prior}\right)}{p\left(D|I_{\rm prior}\right)}
\end{equation}

\noindent the probability $p\left(D|I_{\rm prior}\right)$ is considered
as a normalisation factor, and its determination is avoided, while the
factor $p\left(r',\rho'|I_{prior}\right)$ correspond to the {\it prior}
distribution. The second factor in the equation can be factorised as

\begin{equation}
\label{eq.probdata}
    \begin{split}
      p\left(D_{\rm kin}|r',\rho',D_{\rm rad},I_{\rm prior}\right) & =
      p\left(Gal|r',\rho',D_{\rm rad},I_{\rm prior}\right){\bf \cdot} \\
      &\quad p\left(GCs|r',\rho',D_{\rm rad},I_{\rm prior}\right){\bf \cdot} \\
      &\quad p\left(PNe|r',\rho',D_{\rm rad},I_{\rm prior}\right)
    \end{split}
\end{equation}

\begin{figure}    
\includegraphics[width=85mm]{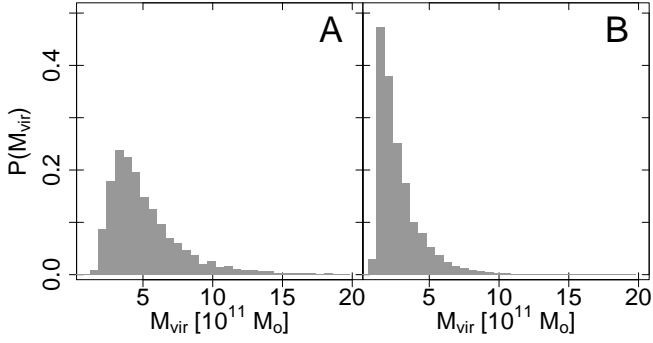}    
\caption{Distribution of virial masses ($M_{\rm vir}$)for dark matter
haloes from the MDPL2 simulation, that results from the first constraint
presented in Section\,\ref{sec.prior}, for central (case A, left panel)
and satellite haloes (case B, right panel).}
\label{dMDM}    
\end{figure}

\noindent where $Gal$ represents the velocity dispersion profile of the
stellar population of the galaxy, and its probability for a specific set
of $r',\rho'$ is obtained through the $\chi^2$ distribution, from the
expression

\begin{equation}
  \chi^2 = \sum_j \left(\frac{{\sigma^{\rm obs}_{\rm LOS,j}} - {\sigma^{\rm pred}_{\rm LOS,j}}}{{e\sigma^{\rm obs}_{\rm LOS,j}}}\right)^2
\end{equation}

\noindent where ${\sigma^{\rm obs}_{\rm LOS,j}}$ represents the measurements
of the velocity dispersion in the line-of-sight, already presented in
Section\,\ref{obs.gmos}, and ${e\sigma^{\rm obs}_{\rm LOS,j}}$ corresponds to
the associated error. In the case of ${\sigma^{pred}_{\rm LOS,j}}$, it
symbolises the value predicted by the model with parameters $r',\rho'$,
and fixed anisotropy $\beta$. The sum is over the measurements at projected
galactocentric distances ($R_{\rm gal}$) larger than 1\,arcmin, for which
rotation fades, and presenting ${e\sigma_j}^{\rm obs}$ smaller than
50\,km\,s$^{-1}$.

The joint probability of the sample of GCs is obtained as the product of
the individual probabilities for each object. These are obtained for each
GC from the convolution of its observational $V_{\rm LOS}$ distribution, and
the predicted one at the $R_{\rm gal}$.
The first one is assumed as a Gaussian centred at the measured $V_{\rm LOS}$
with dispersion equal to the error of the measurement. The latter one
is centred in the mean $V_{\rm LOS}$ of the sample, $685$\,km\,s$^{-1}$,
which agrees with the systemic velocity of the galaxy. It is expressed in
terms of the Gauss-Hermite polynomials \citep{vdm93}, with terms that
characterise asymmetric deviations from the normal distribution vanishing
based on the spherical symmetry,

\begin{equation}
  f(w) = \frac{1}{\sqrt{2\pi}\sigma}{\rm exp}\left(-\frac{1}{2}w^2\right)\left[1+\frac{h_4}{\sqrt{24}}(4w^4-12w^2+3)\right]
\end{equation}

\noindent plus higher-order terms which are assumed negligible in this
approach, $w=(V_{\rm LOS}-V_0)/\sigma_{\rm LOS}$ and $h_4$ is related with the
fourth-order moment of the velocity in the line-of-sight by
$h_4 = (\overline{V_{\rm LOS}^4}/\sigma_{\rm LOS}^4-3)/(8\sqrt{6})$, with
$\sigma_{\rm LOS}$ and $\overline{V_{\rm LOS}^4}$ arising from the Jeans
analysis, for each pair of parameters $r'$, $\rho'$. The same treatment
is applied to the PNe sample, but in this case the mean $V_{\rm LOS}$ from
the sample results $711$\,km\,s$^{-1}$. 

\subsubsection{{\it Prior} distribution}
\label{sec.prior}
The information in $I_{\rm prior}$ is based on two constraints for the
halo parameters. The first one is built from the haloes in the MDPL2
simulation with absolute magnitudes in the $r$ filter consistent with
the brightness of NGC\,3377. The total apparent magnitude in the $r$
band for the galaxy is obtained from the integration of the surface
brightness profile derived by \citet{kra13}, and the extinction
correction from Section \ref{photgal}. It results $r=10.11\pm0.08$\,mag.
From the previous magnitude and the distance already presented in
Section\,\ref{intro} (that implies a distance modulus of 
$m-M= 30.13\pm0.08$\,mag), results the distribution of absolute 
magnitudes ($M_r$), which is used to statistically select haloes 
from the MDPL2 simulation, based on their resulting $M_r$ from the 
HOD method. Two scenarios of {\it prior} distribution are considered, 
in case (A) only central haloes are used to build it, and in case (B) 
only satellite haloes are taken into account.
In Figure\,\ref{dMDM} are shown the distribution of virial masses
($M_{\rm vir}$) built from the haloes in the MDPL2 simulation for both
scenarios (central and satellite haloes), i.e. $p(M_{\rm vir}|I_{\rm prior})$,
that emerges from the first constraint. The two panels present the cases
(A) and (B), achieving the 95-percentile at $M_{\rm vir}$ values of $10^{12}$ and
$6\times10^{11}\,{\rm M_{\odot}}$, respectively. The restriction to satellites
favours less massive haloes due to the differences in the luminosity
functions for central and satellite galaxies from \citet{lan16}. This
is expected as a consequence of the mass loss in subhaloes beyond their
accretion, caused by dynamical friction, tidal stripping and tidal
heating processes \citep[e.g.][]{gan10}. In this regard, \citet{cor18}
indicated for haloes less massive than $\approx 10^{12}\,{\rm M_{\odot}}$,
that central galaxies inhabit more massive haloes than satellites at
fixed stellar mass.

\begin{figure}    
\includegraphics[width=85mm]{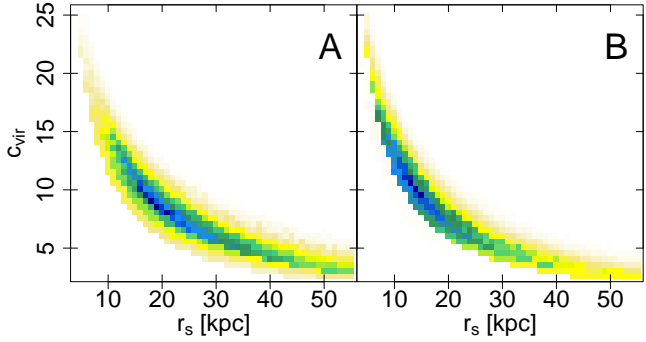}    
\caption{{\it Prior} distribution of the parameters describing the NFW
profile, derived from both constraints in Section\,\ref{sec.prior}, for
central haloes (case A, left panel) and satellite ones (case B, right
panel). The colour gradient, from light yellow to dark blue, represents
sets of parameters with increasing probability.}
\label{fig.prior}    
\end{figure}

The second constraint comes from \citet{cap13a,cap13b}, who found for
NGC\,3377 dynamical and stellar mass-to-light ratios ($M/L_r$) corresponding
to a dark matter fraction of $\approx6\%$ up to $1\,R_{\rm eff}$, with errors
for the fitted $M/L_r$ of $\approx6\%$. Considering that both constraints
apply on $M_{\rm vir}$, it results

\begin{equation}
\label{eq.prior}
  p(r',\rho'|I_{\rm prior}) = p(r',\rho'|M_{\rm vir})p(M_{\rm vir}|I_{\rm prior})
\end{equation}

\begin{figure*}    
\includegraphics[width=55mm]{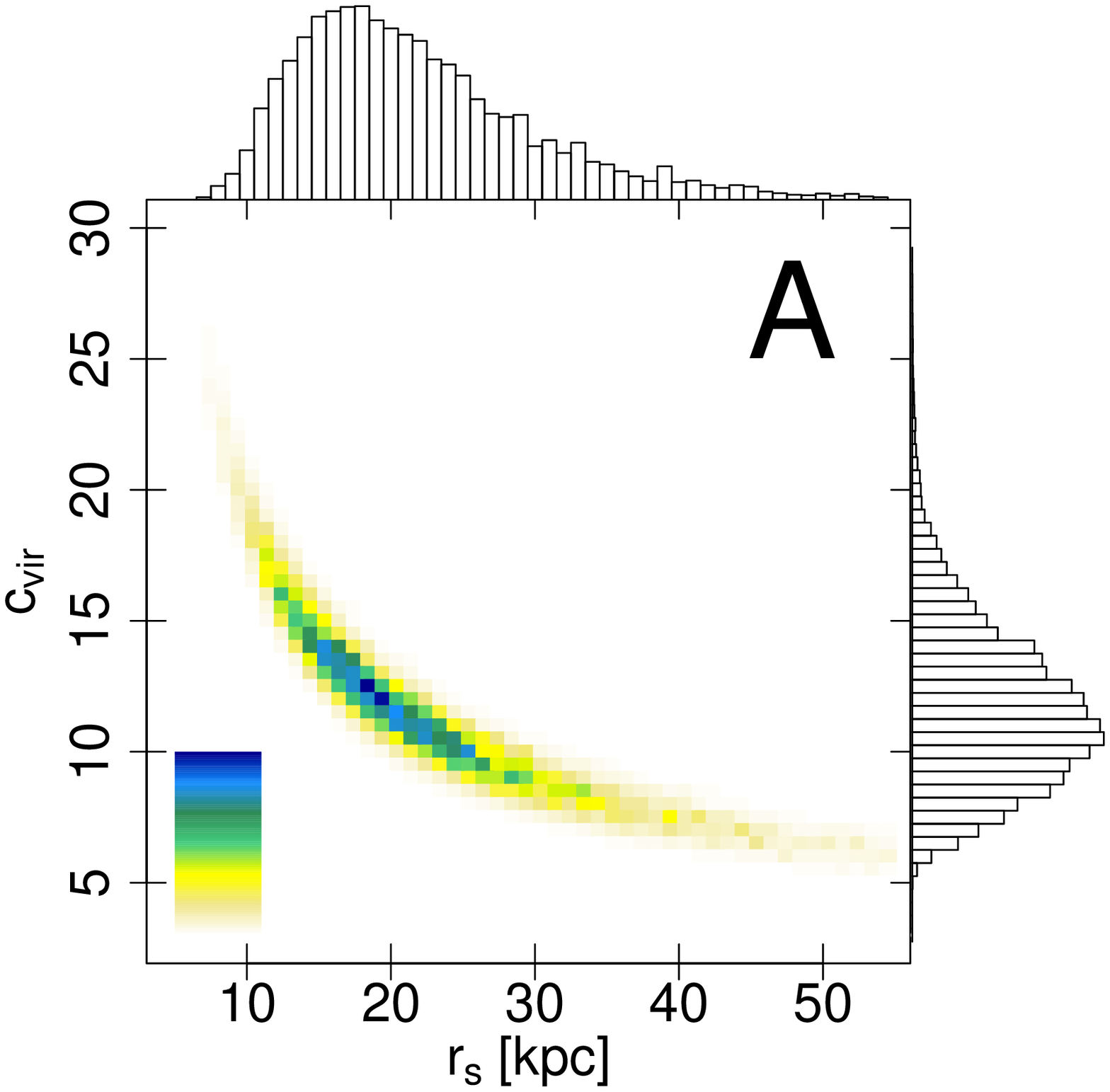}    
\includegraphics[width=55mm]{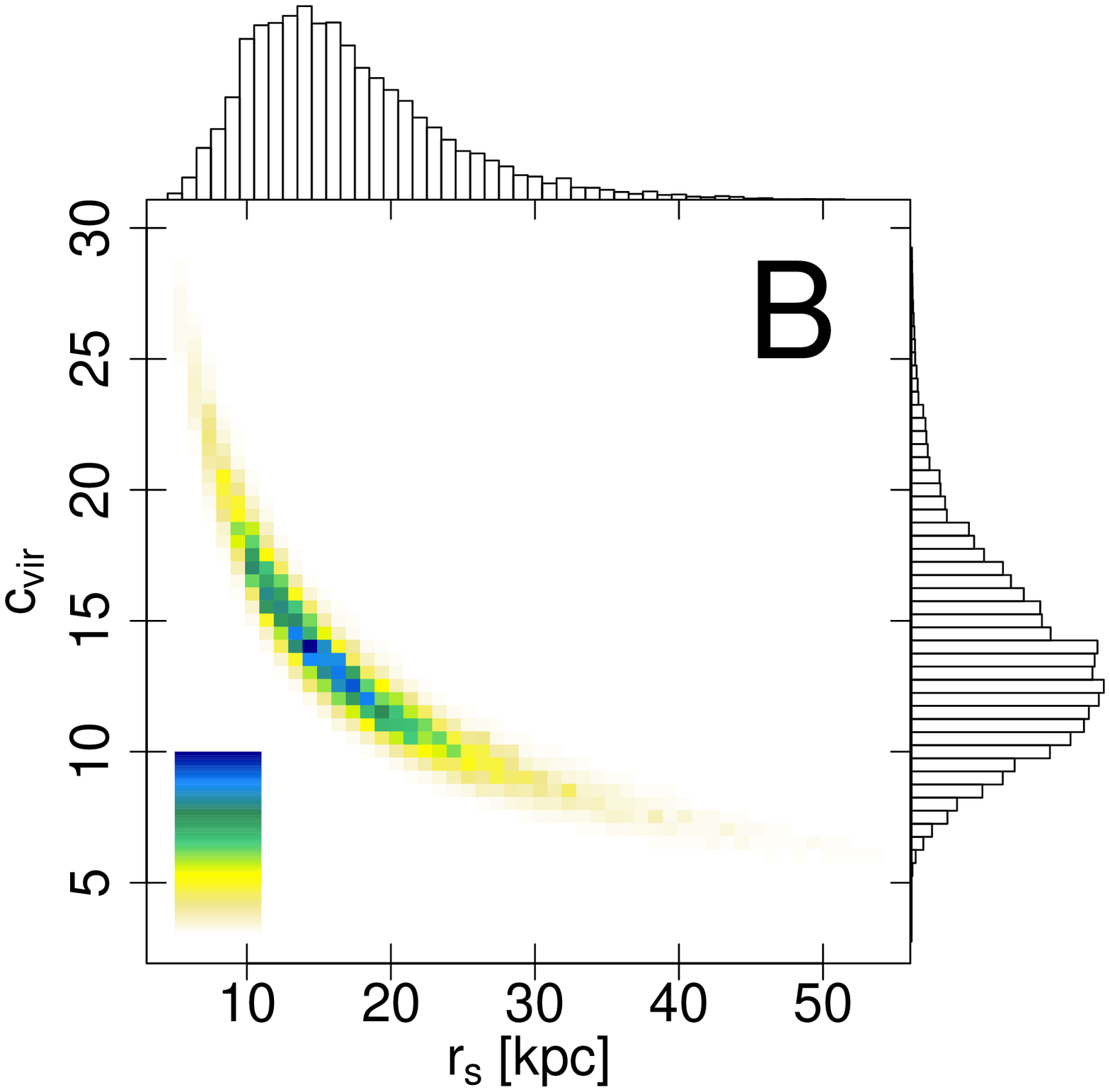}  
\includegraphics[width=55mm]{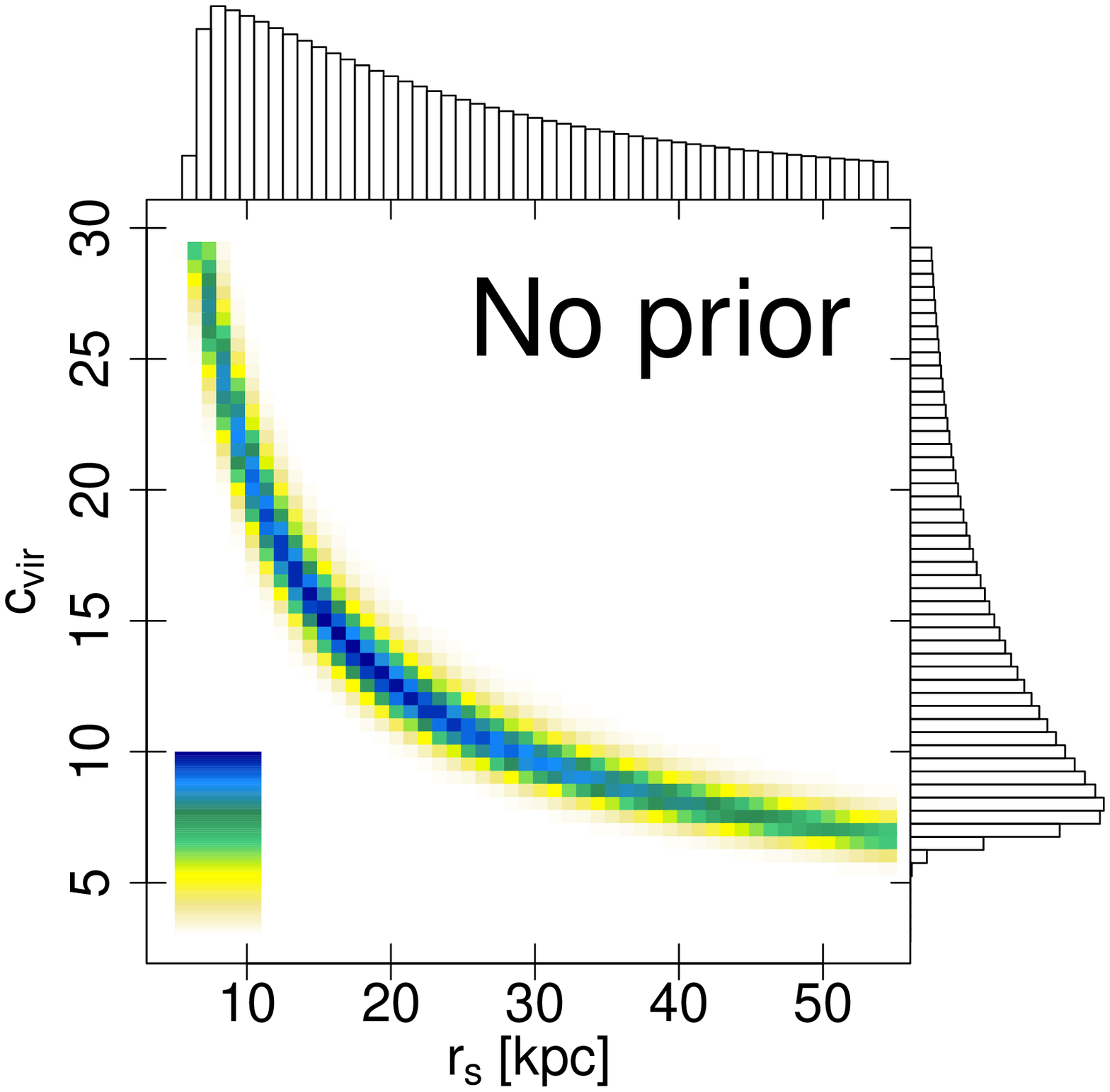}\\  
\caption{Colour map of the NFW profiles for the joint probability of
the stellar population, GCs and PNe, when different anisotropy 
parameters are assumed for each tracer population ($\beta_{\rm GCs}=-0.5$, 
$\beta_{\rm stel}=0$ and $\beta_{\rm PNe}=0.5$).
The NFW profile is described in terms of the concentration parameter
($c_{\rm vir}$) and its characteristic radius ($r_{\rm s}$). The colour
gradient evolves from light yellow to dark blue, representing
increasing joint probabilities. The three panels differ in the {\it prior}
distribution, built from central haloes (case A, left panel), satellite
haloes (case B, middle panel), and no {\it prior} (right panel). The side
histograms correspond to the marginal distributions for $c_{\rm vir}$ and
$r_{\rm s}$.
}
\label{fig.iso}    
\end{figure*}

\section{Results}

In this Section are presented the results obtained from the analysis
already described in Section\,\ref{sec.model}, based on the observational
dataset gather in this paper, and presented in Section\,\ref{sec.obs}.
A cuspy mass distribution is assumed, applying both the NFW and 
Einasto models. Although core-like halo distributions accurately represent 
the mass profile in dwarf galaxies, it is not clear whether the haloes from 
galaxies in the mass range of NGC\,3377 significantly deviate from the inner 
slopes represented by the assumed profiles \citep[e.g.][]{del16,oh15}. Also 
the scenarios in which NGC\,3377 behaves as a central or satellite halo are 
considered to define the {\it prior} distribution (see Section\,\ref{sec.prior} 
for further details), as well as several cases of constant anisotropy ($\beta$).

\subsection{The NFW profile}
\label{sec.nfw}

The NFW profile \citep{nav97}, commonly used for modelling dark matter
haloes, can be described by a characteristic radius ($r_{\rm s}$) and density
($\rho_{\rm s}$). The mass profile in terms of these two parameters is

\begin{equation}
  M_{\rm NFW}(r) = 4\pi\rho_{\rm s} r^3_{\rm s}\left[ln\left(1+r/r_{\rm s}\right)-\frac{r/r_{\rm s}}{1+r/r_{\rm s}}\right]
\end{equation}

Following the definitions from \citet{bul01}, the concentration parameter
is the ratio between the virial ($R_{\rm vir}$) and characteristic radii,
$c_{\rm vir}=R_{\rm vir}/r_{\rm s}$, with $R_{\rm vir}$ the radius at which the mean
density within it equals to $\Delta_{\rm vir}(\Omega_{\rm m})$ times the mean
matter density ($\rho_{\rm m}$). For $\Omega_{\rm m}=0.307$ and the approximation
from \citet{bry98}, it results $\Delta_{\rm vir} \approx 333$. From the
definition of the density parameter $\Omega_{\rm m}$, it results 
$\rho_{\rm m}= \Omega_{\rm m} \rho_{\rm crit}$, where the last factor corresponds
to the critical density for a flat Universe, $\rho_{\rm crit}=3H^2_0/(8\pi G)$.
Then, it leads to a relation involving $\rho_{\rm s}$ and $c_{\rm vir}$

\begin{equation}
  \rho_{\rm s} = 100 \frac{H^2_0}{8\pi G}\frac{c^3_{\rm vir}}{ln(1+c_{\rm vir})-c_{\rm vir}/(1+c_{\rm vir})}
\end{equation}

\noindent with $H_0=67.77$\,km\,s$^{-1}$\,Mpc$^{-1}$ and $G$ the gravitational
constant. Then, the expression for the virial mass of a NFW halo
directly arises, from these equations and the definition of
$c_{\rm vir}$, $M_{\rm vir} = 51.1 H^2_0 (c_{\rm vir} r_{\rm s})^3/G$

Hereafter, the parameters $r_{\rm s}$ and $c_{\rm vir}$ describes the
NFW profile, instead of $\rho_{\rm s}$. In Figure\,\ref{fig.prior} is
shown the {\it prior} distribution derived from both restrictions
on the $M_{\rm vir}$, presented in Section\,\ref{sec.prior}, and applied
to the NFW parameters. The colour gradient ranges from light yellow
to dark blue towards pairs of $(r_{\rm s},c_{\rm vir})$ with larger
probability. The panels correspond to central and satellite haloes.

\begin{table*}  
\begin{minipage}{160mm}   
\begin{center}   
\caption{Solutions for the kinematical analysis from the three sets of
tracers, i.e. stellar population, GCs and PNe. For each parameter of 
the density distribution, the mode, mean and dispersion are indicated in 
three consecutive columns. Two anisotropy options are assumed, in the 
first one the isotropic case is applied for the three sets of tracers, 
and the second option considered different anisotropy parameters for each 
set (see text in Section\,\ref{sec.nfw}). For each anisotropy option, the 
solutions listed in consecutive rows differ in the {\it prior} distribution, 
built from central haloes (case A), satellite haloes (case B), or no 
{\it prior} distribution (third row). In these latter cases, the large 
dispersion in the distribution of joint probabilities prevent us from 
indicating the modes of the parameters. The distribution parameters are
$r_{\rm {\rm s}}$ and $c_{\rm vir}$ for the NFW profile, and $r_{-2}$ and
$\rho_{-2}$ in the case of the Einasto profile. In this latter case, the 
third parameter is fixed at $n=8$ (further details in the text). For both
density distributions, also the $M_{\rm vir}$ estimators are shown in the 
last three columns.}
\label{dNFW}   
\setlength{\tabcolsep}{4pt}
\begin{tabular}{@{}cccccccccc@{}}
\hline 
\hline
\multicolumn{10}{c}{NFW profile}\\
\hline
\multicolumn{1}{@{}c}{Prior}&\multicolumn{3}{c}{$r_{\rm s} {\rm [kpc]}$}&\multicolumn{3}{c}{$c_{\rm vir}$}&\multicolumn{3}{c}{$M_{\rm vir} {\rm [10^{11}\,M_{\odot}]}$}\\   
\hline   
\multicolumn{10}{c}{$\beta=0.0$}\\
\hline   
A & $18\pm2.7$ & $21.4\pm2.5$ & $ 8.3\pm1.0$ & $12.5\pm1.5$ & $12.2\pm1.1$ & $3.1\pm0.2$ & $4.1\pm1.1$ & $6.1\pm1.1$ & $2.6\pm0.5$\\
B & $15\pm2.5$ & $18.0\pm2.3$ & $ 7.1\pm1.0$ & $13.5\pm1.5$ & $13.5\pm1.0$ & $3.5\pm0.3$ & $3.6\pm0.9$ & $4.9\pm1.2$ & $2.1\pm0.5$\\
- & $   -    $ & $24.1\pm5.1$ & $13.2\pm1.5$ & $    -     $ & $13.7\pm2.3$ & $5.7\pm0.7$ & $    -    $ & $6.3\pm1.9$ & $4.2\pm1.3$\\
\hline
\multicolumn{10}{c}{$\beta_{\rm GCs}=-0.5$, $\beta_{\rm stel}=0.0$, $\beta_{\rm PNe}=0.5$}\\
\hline   
A & $18\pm3.1$ & $22.1\pm2.5$ & $ 8.7\pm1.1$ & $12.5\pm1.7$ & $12.4\pm1.2$ & $3.1\pm0.2$ & $4.1\pm1.0$ & $5.8\pm1.1$ & $2.4\pm0.5$\\
B & $14\pm2.5$ & $17.0\pm2.3$ & $ 6.9\pm1.0$ & $14.0\pm1.4$ & $13.5\pm1.1$ & $3.5\pm0.3$ & $3.6\pm0.9$ & $4.6\pm1.1$ & $2.0\pm0.5$\\
- & $   -    $ & $24.1\pm5.4$ & $13.2\pm1.3$ & $    -     $ & $13.5\pm2.6$ & $5.6\pm0.7$ & $    -    $ & $6.3\pm1.7$ & $4.1\pm1.2$\\
\hline   
\hline 
\multicolumn{10}{c}{Einasto profile}\\
\hline   
\multicolumn{1}{@{}c}{Prior}&\multicolumn{3}{c}{$r_{-2} {\rm [kpc]}$}&\multicolumn{3}{c}{$\rho_{-2} {\rm [10^{-3}\,M_{\odot}\,pc^{-3}]}$}&\multicolumn{3}{c}{$M_{\rm vir} {\rm [10^{11}\,M_{\odot}]}$}\\   
\hline   
\multicolumn{10}{c}{$\beta=0.0$}\\
\hline   
A & $17\pm4.2$ & $19.8\pm2.8$ & $9.0\pm0.7$ & $1.0\pm0.2$ & $1.3\pm0.2$ & $0.9\pm0.1$ & $4.4\pm2.2$ & $5.6\pm2.0$ & $2.1\pm0.7$\\ 
B & $12\pm3.5$ & $16.6\pm2.7$ & $7.5\pm0.8$ & $1.8\pm0.4$ & $1.7\pm0.2$ & $1.1\pm0.1$ & $3.8\pm1.6$ & $4.7\pm1.9$ & $1.7\pm0.7$\\
- & $    -   $ & $17.7\pm1.8$ & $9.3\pm1.4$ & $    -    $ & $2.2\pm0.3$ & $1.4\pm0.1$ & $    -    $ & $7.1\pm2.1$ & $2.7\pm0.5$\\
\hline   
\multicolumn{10}{c}{$\beta_{\rm GCs}=-0.5$, $\beta_{\rm stel}=0.0$, $\beta_{\rm PNe}=0.5$}\\
\hline   
A & $22\pm4.1$ & $19.5\pm2.8$ & $8.9\pm0.7$ & $0.6\pm0.2$ & $1.2\pm0.2$ & $0.9\pm0.1$ & $4.3\pm1.9$ & $5.1\pm2.0$ & $1.9\pm0.8$\\
B & $12\pm3.4$ & $15.8\pm2.8$ & $7.5\pm0.8$ & $1.8\pm0.3$ & $1.5\pm0.2$ & $1.1\pm0.1$ & $3.3\pm1.5$ & $4.2\pm1.9$ & $1.5\pm0.7$\\
- & $    -   $ & $17.8\pm2.0$ & $9.9\pm1.5$ & $    -    $ & $2.1\pm0.4$ & $1.4\pm0.1$ & $    -    $ & $6.2\pm2.0$ & $2.5\pm0.5$\\
\hline
\hline

\end{tabular}    
\end{center}    
\end{minipage}   
\end{table*}   

Following Equation\,\ref{eq.bayes}, the joint probability for each
pair of parameters $(r_{\rm s},c_{\rm vir})$ results from the product of
Equations\,\ref{eq.probdata} and \ref{eq.prior}, with the latter one
changing for cases (A) and (B), and the former equation depending on
the orbital anisotropy ($\beta$).
Although more complex treatments for this parameter are derived in
the literature \citep[e.g.][]{mam05}, in this paper only solutions with
constant anisotropy are considered. The degeneracy between solutions
with different values of $\beta$ might be partially resolved analysing
the kurtosis of the velocity distribution for the stellar component
\citep[][]{nap09,sal12} and the estimators from Section\,\ref{sec.kin}.
The appendix\,C in \citet{coc09} shows the distribution of $h4$ parameter
for the major and minor axis of NGC\,3377, the values are consistent
with kurtosis $\approx 0$ between 60 and 120\,arcsec, but the data points
are scarce and the uncertainties are large. The $\beta$ parameter up to
the effective radius for a sample of elongated ellipticals from the
SAURON project \citep{cap07} span from $\approx-0.2$ to $\approx+0.2$,
with NGC\,3377 being nearly isotropic (see their Table\,2).

The first rows in Table\,\ref{dNFW}
shows the mode, mean and dispersion for the distributions of $r_{\rm s}$
and $c_{\rm vir})$, for the isotropic case ($\beta=0$) and the three options 
regarding the {\it prior} distribution (i.e., central haloes, satellite
haloes, and no {\it prior}). Also the parameters for the distribution of 
$M_{\rm vir}$ are indicated in the last three columns. In the cases of no 
{\it prior} distribution, the large dispersion in the distribution of joint
probabilities prevent us from indicating the modes of the parameters.
The uncertainties come from the parameters dispersion from 100 Monte-Carlo 
realisations on artificial samples, generated to represent the observational 
data set and its uncertainties (see Section\,\ref{sec.test}). Also a 
set of mild constant anisotropies, ranging from $\beta=-0.5$ to $\beta=0.5$
have been used. These results are consistent, with modes and means pointing
to more massive (modes of $M_{\rm vir}$ ranging from $3.4\pm1.0$ to $6.1\pm1.2
\times 10^{11}\,{\rm M_{\odot}}$ for central haloes and from $2.9\pm0.9$ to 
$5.4\pm0.9\times 10^{11}\,{\rm M_{\odot}}$ for satellites) and less 
concentrated haloes (modes of $c_{\rm vir}$ ranging from $11\pm2$ to $13\pm1.5$ 
for central haloes). The {\it prior} distribution restricted to satellite 
haloes favours less massive haloes than that for central ones, reflecting the 
differences that arise from the first constraint in Section\,\ref{sec.prior}.
The last fitting option for the NFW profile in Table\,\ref{dNFW} 
corresponds to solutions that account for different anisotropy parameters for 
each kinematical tracer. For the stellar population is assumed the isotropic 
case ($\beta_{\rm stel}=0.0$), which does not contradict the $h_4$
measurements at Appendix\,C from \citet{coc09} and Table\,2 in \citet{cap07}. 
The GCs kinematical data is fitted considering $\beta_{\rm GCs}=-0.5$, and 
$\beta_{\rm PNe}=0.5$ is assumed for the PNe. These values are based on the 
kurtosis estimators from Section\,\ref{sec.kin}. In Figure\,\ref{fig.iso} 
are shown the distributions of the parameters $(r_{\rm s},c_{\rm vir})$ for 
this latter case, with the
{\it prior} distributions corresponding to central haloes (case A, left
panel), satellite haloes (case B, middle panel), and without {\it prior}
distribution (right panel). The grid of parameters is built with steps
$\Delta c_{\rm vir}=0.5$ and $\Delta r_{\rm s}=1$\,kpc, and represents a
wide range of $M_{\rm vir}$. The colour gradient ranges from light yellow
to dark blue towards halo parameters $(r_{\rm s},c_{\rm vir})$ with larger
joint probability, according to both the fit of the dynamical tracers
and the {\it prior} distributions (in the cases that applies). The side
histograms in each case correspond to the marginal distributions for
$c_{\rm vir}$ and $r_{\rm s}$.

\subsection{The Einasto profile}

\begin{figure*}    
\includegraphics[width=55mm]{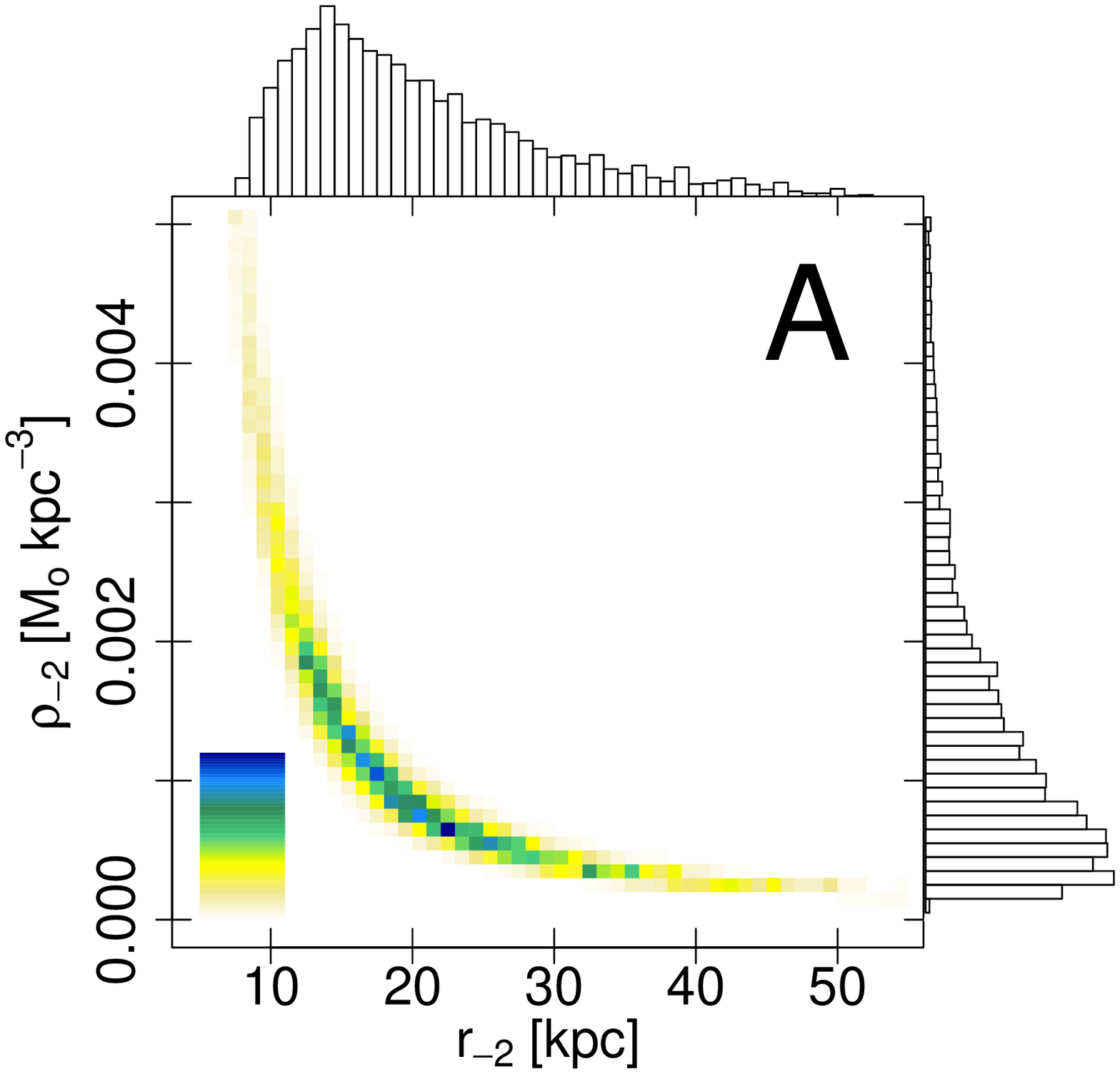}    
\includegraphics[width=55mm]{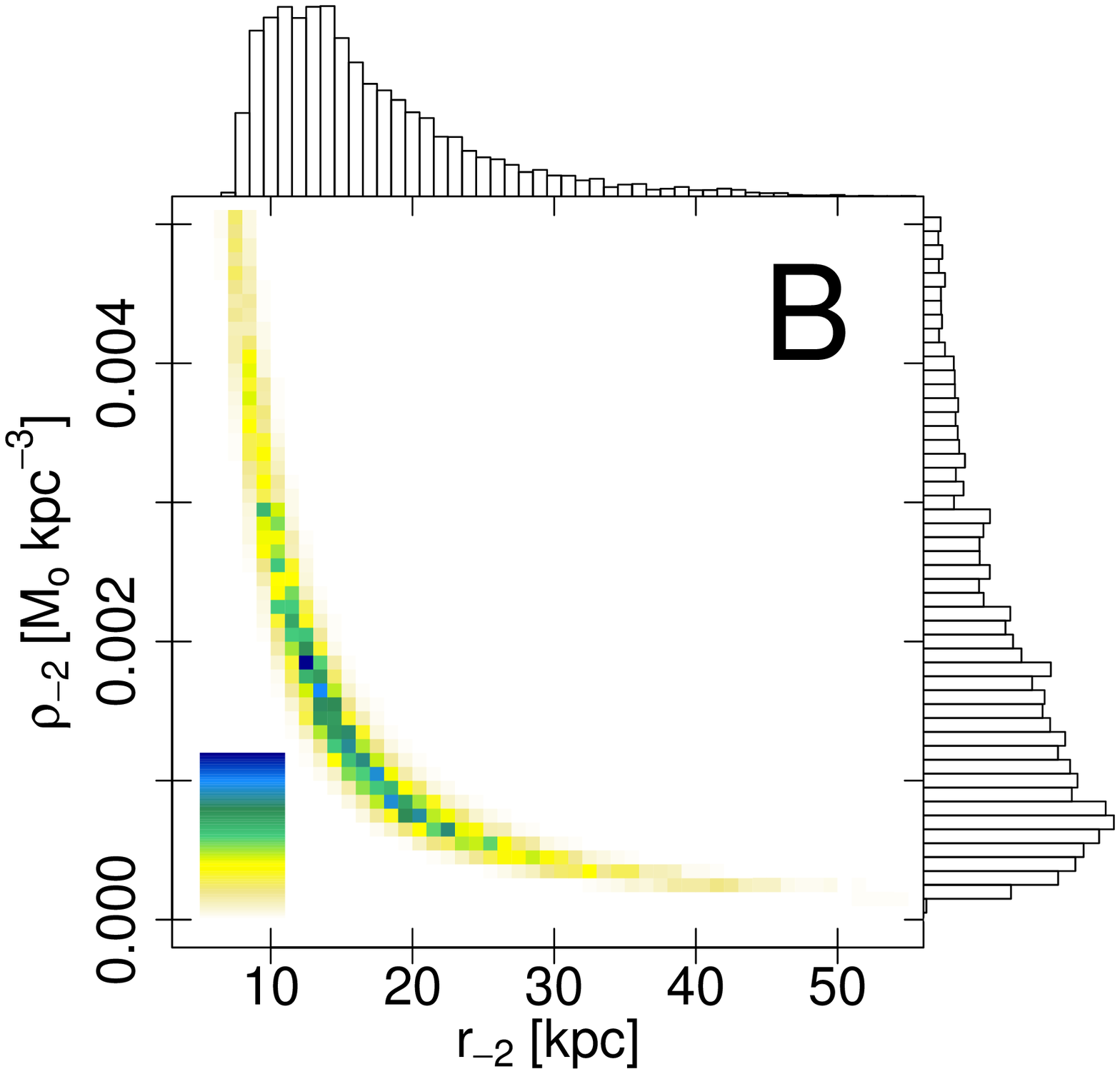}  
\includegraphics[width=55mm]{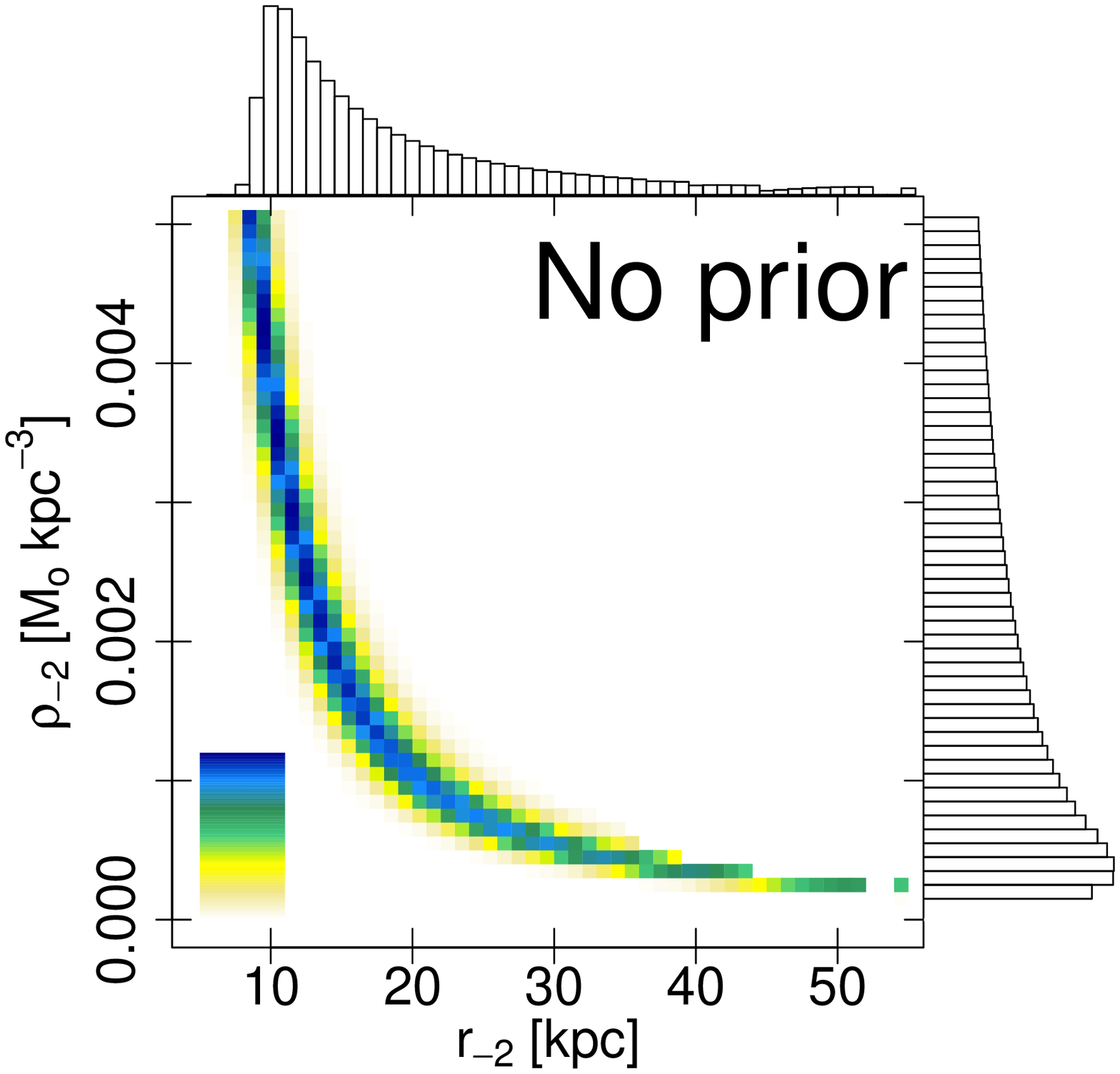}\\  
\caption{Colour map of the Einasto profiles for the joint probability of
the stellar population, GCs and PNe, when different anisotropy 
parameters are assumed for each tracer population ($\beta_{\rm GCs}=-0.5$, 
$\beta_{\rm stel}=0$ and $\beta_{\rm PNe}=0.5$).
The mass profile is described in terms of the characteristic radius
($r_{-2}$) and the mass density at this value ($\rho_{-2}$). The colour
gradient represents increasing probability, from yellow to dark blue.
The three panels differ in the {\it prior} distribution, built from
central haloes (case A, left panel), satellite haloes (case B, middle
panel), and no {\it prior} (right panel).}
\label{fig.isoein}    
\end{figure*}

The Einasto profile \citep{ein65} has been useful to describe
the density of dark matter haloes in numerical simulations, and
its properties have been extensively analysed in the literature
\citep[e.g.][and references there in]{ret12}. It possesses a
power-law logarithmic slope, leading to the expression:

\begin{equation}
  \rho_{\rm Ein}(r) = \rho_{-2} {\rm exp}\left\{ -2n\left[\left(\frac{r}{r_{-2}}\right)^{1/n}-1\right]\right\}
\end{equation}

\noindent with $r_{-2}$ acting as a characteristic radius where the
density profile behaves as a power-law with exponent $-2$, $\rho_{-2}$
represents the density at this radius, and $n$ is a positive number
reflecting the steepness of the power-law. The mass profile in terms
of these parameters is

\begin{equation}
  \begin{split}
    M_{\rm Ein}(r) & = M_* \gamma\left(3n,2n \sqrt[n]{\frac{r}{r_{-2}}}\right), \\
    M_* & =   4\pi \rho_{-2} r^3_{-2} n\frac{e^{2n}}{(2n)^{3n}} \\
  \end{split}
\end{equation}

\begin{figure*}    
\includegraphics[width=92mm]{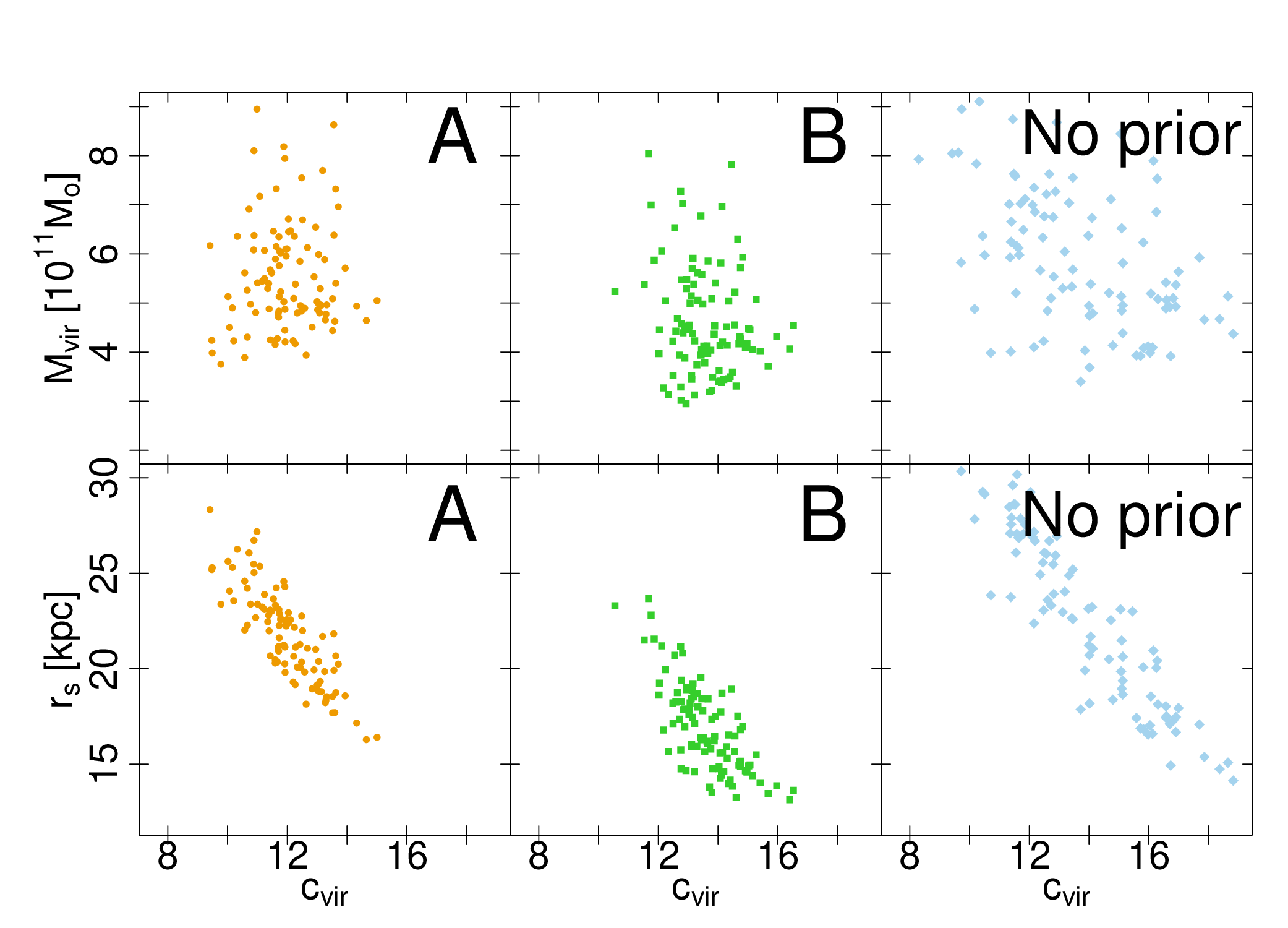}    
\includegraphics[width=73mm]{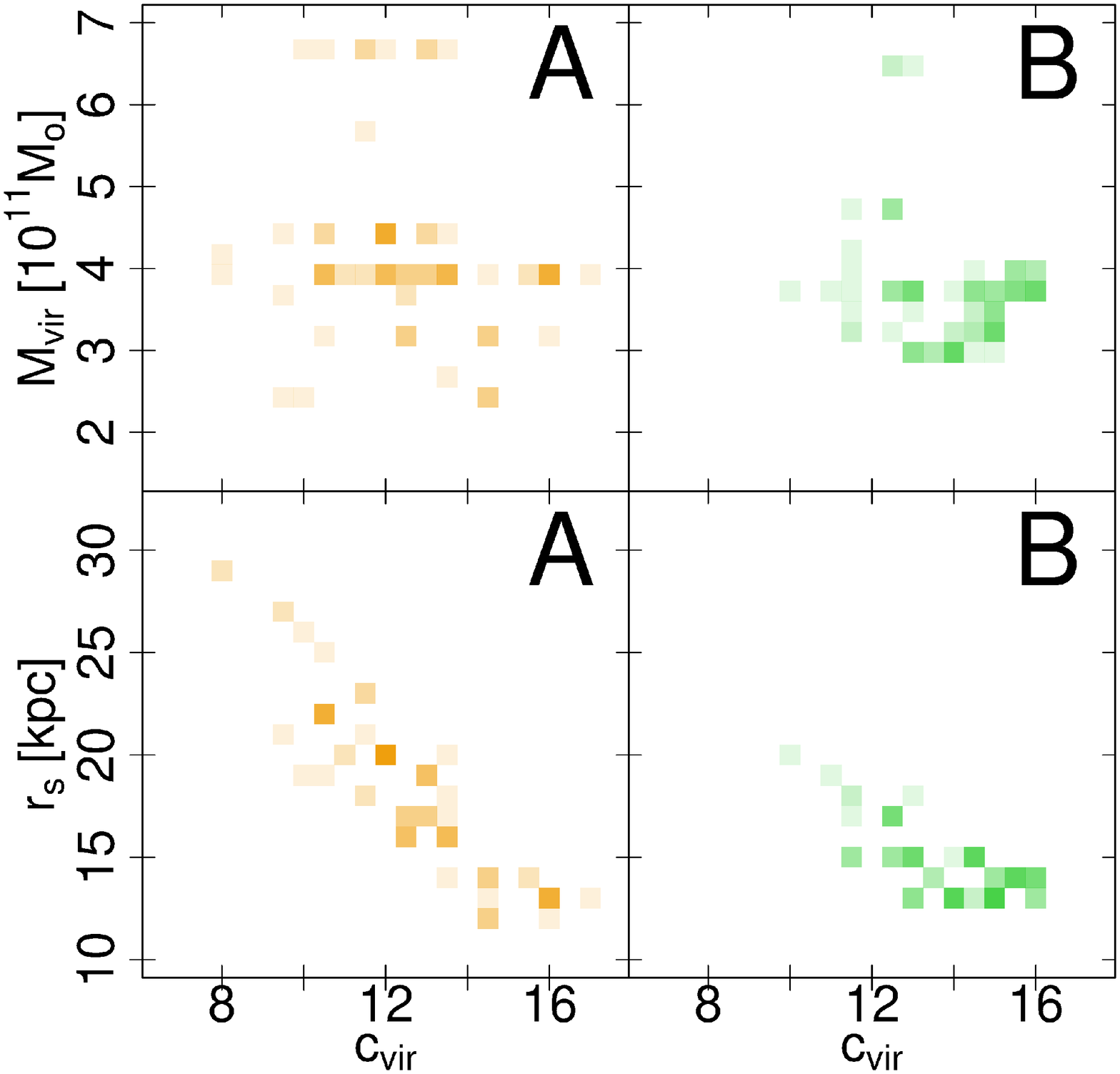}    
\caption{{\bf Left panels:} Means of the distributions of parameters
for the NFW profile, obtained in the same manner than those from the
observational dataset, for 100 Monte-Carlo realisations when 
different anisotropy parameters are assumed for each tracer population 
($\beta_{\rm GCs}=-0.5$, $\beta_{\rm stel}=0$ and $\beta_{\rm PNe}=0.5$). 
The three options differ in the {\it prior} distribution, assuming 
central haloes (case A), satellite haloes (case B), and no {\it prior} 
distribution. {\bf Right panels:} Modes of the distributions of parameters 
from 100 Monte-Carlo realisations, obtained for the same anisotropy 
case and the NFW profile. The colour gradients, increasing from light 
to dark, represents more repeated sets of parameters. Only cases A and B 
are presented this time.}
\label{fig.isomodel}    
\end{figure*}

\noindent with $\gamma(a,x)$ corresponding to the lower incomplete
gamma function. Its $R_{\rm vir}$ is derived from this latter expression
and its definition, looking for the value $R_{\rm vir}$ that leads to
a mean density in the halo that equals $333$ times the mean matter
density ($\rho_{\rm m}$, see Section\,\ref{sec.nfw}). Then, evaluating
the previous equation at $R_{\rm vir}$ leads to the virial mass, $M_{\rm vir}$.
Defining $\alpha = n^{-1}$, \citet{gao08} found that
$\alpha$ is related with the halo mass and the redshift through the
rms fluctuation of the density field $\sigma(M,z)$, varying very
little for present day haloes with masses below $10^{13}\,{\rm M_{\odot}}$.
A similar relation was fitted by \citet{kly16} from the suite of
Multidark cosmological simulations. From this latter relation, it
is assumed $\alpha = 0.125$ for the range of virial masses expected
for NGC\,3377, resulting in $n= 8$. Then, the profile is described
by only two free parameters, $r_{-2}$, and $\rho_{-2}$, for whom the
{\it prior} distributions are built from the two constraints presented
in Section\,\ref{sec.prior}, in the same manner than those for the NFW
profile.

The lower half of 
Table\,\ref{dNFW} lists the mode, mean and dispersion for the
free parameters of the Einasto profile, and for the $M_{\rm vir}$
distribution, when the {\it prior} distribution is derived from
the central haloes population (case A) or from the satellite haloes
one (case B), and also the results for no {\it prior} distribution
at all (third row). In these latter cases, the large dispersion in
the distribution of joint probabilities prevent us from indicating
the modes of the parameters. The uncertainties are obtained
from 100 Monte-Carlo realisations that reproduced the sizes and errors
of the observational data set (see Section\,\ref{sec.test}). 
In the first option, it is shown the isotropic case for the three 
tracer populations, although a larger set of anisotropy parameters 
(from $\beta=-0.5$ to 0.5) are considered. Increasing $\beta$ values 
result in more massive haloes (the mode of $M_{\rm vir}$ ranges 
from $4.3\pm1.7$ to $7.8\pm2.6\,{\rm M_{\odot}}$ for central haloes, 
and from $3.3\pm1.5$ to $6.8\pm2.4\,{\rm M_{\odot}}$ for satellites), 
with density profiles that steepen at larger radii (the mode of 
$r_{-2}$ ranges from $20\pm4.0$ to $25\pm4.5$\,kpc for central haloes,
and from $11\pm3.4$ to $22\pm4$\,kpc for satellites). This behaviour
is similar to that derived from NFW profiles in the previous section.
Also a solution with different choices of $\beta$ parameters for each
tracer population (i.e. $\beta_{\rm stel}=0.0$ for the stellar 
population, $\beta_{\rm GCs}=-0.5$ for the GCs, and $\beta_{\rm PNe}=0.5$ 
for the PNe) is applied, analogue to the NFW analysis. As expected,
in all cases the {\it prior} distribution derived from satellite 
haloes leads to lighter haloes. The distributions of the parameters 
$(r_{-2},\rho_{-2})$ for the latter anisotropy case are shown in 
Figure\,\ref{fig.isoein}. The three panels represent the results when 
the {\it prior} distributions correspond to cases (A) and (B),
as well as the results whether no {\it prior} distribution is
applied. For this profile, the steps of the grid of parameters are
$\Delta r_{-2}=1$\,kpc and $\Delta \rho_{-2}=10^{-4}\,{\rm M_{\odot}\,pc^{-3}}$.
The colour gradient ranges from light yellow to dark blue towards
halo parameters $(r_{-2},\rho_{-2})$ with larger joint probability.
The side histograms in each case correspond to the marginal
distributions for $r_{-2}$ and $\rho_{-2}$.

\subsection{Testing the procedure}
\label{sec.test}

In order to test the accuracy of the procedure, 100 Monte-Carlo
realisations are generated for each option of mass profile (i.e.,
NFW or Einasto), $\beta$ parameter, and {\it prior} distribution.
In each case, the parameters that describe the respective mass
profile of the haloes are assumed as those listed in
Table\,\ref{dNFW}. The size and galactocentric ranges of the 
artificial samples match those from the observational ones. Random 
velocity uncertainties are added to the simulated $V_{\rm LOS}$, 
following the percentile distribution of the velocity errors from 
the observational data set.

Once the artificial sample is generated, the procedure described
in Section\,\ref{sec.model} is applied in the same manner than for
the observational data set. In each step, the mode, mean and
dispersion for the distribution of parameters of the mass profile,
and the $M_{\rm vir}$ distribution are recorded. The results show a
good agreement with the underlying set of parameters. In the left
panels of Figure\,\ref{fig.isomodel}, the mean for the distributions
of parameters in the case of different anisotropy parameter for 
each tracer population are plotted for the NFW profile,
and the three options of {\it prior} distribution. When {\it prior}
distributions are avoided, the 100 realisations leads to more dispersed
results. After rejecting outliers, the 90-percentile of the mean values
in cases (A) and (B) differs from the original parameters in less than
4\,kpc for $r_{\rm s}$, 1.6 for $c_{\rm vir}$, and
$1.7 \times 10^{11}$ ${\rm M_{\odot}}$ for $M_{\rm vir}$, if NFW profiles
are assumed.
For the Einasto profiles, the 90-percentiles of the mean values are
achieved at 5\,kpc for $r_{-2}$, $3.8 \times 10^{-4}\,{\rm M_{\odot}\,pc^{-3}}$,
and $3 \times 10^{11}\,{\rm M_{\odot}}$ for $M_{\rm vir}$. In the right panels
of Figure\,\ref{fig.isomodel} are plotted the distributions of modes for
the NFW profile and the case of different anisotropy parameter for 
each tracer population. The colour gradients, increasing
from light to dark, represent more repeated sets of parameters. Only cases
A and B are showed this time. In comparison with the means of the parameters,
the modes are typically more disperse, due to discreteness effects from the
grid.
For each parameter, the dispersion from these 100 realisations are used to
estimate the uncertainties of the results listed in Table\,\ref{dNFW}.

\section{Comparison with the literature}
\label{sec.comp}

Regarding the M\,96 subgroup, it seems to be a moderately massive group of
galaxies, with $M_{\rm vir}\approx 7 \times 10^{12}\,{\rm M_{\odot}}$
\citep{kar04b}, which implies a virial radius of $\approx 500$\,kpc for a
NFW profile. The  mean harmonic distance for the galaxies is $\approx 190$
kpc, similar to the most probable $R_{\rm vir}$ for NGC\,3377.
From the radial velocities measured for a small sample of GCs, \citet{ber06}
obtained for NGC\,3379, the other moderately bright elliptical, a velocity
dispersion which is consistent with a halo presenting
$M_{\rm vir} \approx 6 \times 10^{11}\,{\rm M_{\odot}}$, which implies
$R_{\rm vir} \approx 220$\,kpc. The
projected distance from NGC\,3379 and M\,96 to NGC\,3377 are $\approx 1.4$\,deg
and $\approx 2.2$\,deg, respectively, corresponding to $\approx 260$\,kpc and
$\approx 400$\,kpc at NGC\,3377 distance. From the mean distances calculated
by NED, the difference in distances between NGC\,3377 and these two galaxies
are $\approx 100$\,kpc and $\approx 300$\,kpc for NGC\,3379 and M\,96,
respectively. \citet{tul13} limited the distance estimators to surface
brightness fluctuations, Cepheids and tip of the red-giant branch, leading
to differences in distance of $\approx 600$ and $\approx 100$\,kpc for
NGC\,3379 and M\,96, respectively, which are lower than the fiducial 10
per-cent uncertainty for distance estimators. Although it is possible that
NGC\,3377 behaves as a main halo, the most probable scenario is that it
belongs to the M\,96 subgroup. Then, the solutions that correspond to case 
(B) in Table\,\ref{dNFW} are preferred. From these,
the anisotropy parameters $\beta\pm0.5$ hardly provide compatible results
with the observational constraints for both GCs and PNe at the same time.
For the remaining cases, the isotropic and that with different anisotropies
for each kinematical tracers, the most probable haloes from the NFW profile
are similar, and the weighted mean of their most probable virial masses
leads to $(3.6\pm0.6) \times 10^{11}\,{\rm M_{\odot}}$. Although the modes
of the $M_{\rm vir}$ distributions for both anisotropy parameters differ
in the case of the Einasto profile, their variation are enclosed by the
uncertainties. The weighted mean results $(3.5\pm1.1) \times 10^{11}
{\rm M_{\odot}}$, in agreement with that derived from the NFW profile.

From Chandra observations, \citet{kim15} measure for the hot gas in
NGC\,3377 a temperature of ${T_{\rm GAS}} \approx 0.2$\,keV and a luminosity
in the range $0.3-8$\,keV of ${L_{\rm X,GAS}} \approx 10^{38}\,{\rm erg s^{-1}}$.
The relation derived by \citet{for17} for the ${L_{\rm X,GAS}}$ leads to a
total mass up to five effective radii of ${M_{\rm tot} (<5\,R_{\rm eff})}$
$\approx (0.5-1)\times10^{11}\,{\rm M_{\odot}}$. Similar results are obtained
from the scaling relations fitted by \cite{bab18} for ${L_{\rm X,GAS}}$ and
${T_{\rm GAS}}$. If the most probable parameters from Table\,\ref{dNFW}
are used to calculate ${M_{\rm tot} (<5\,R_{\rm eff})}$, the results agrees 
with the ranges derived from the X-ray observations. These also agrees with 
the estimations up to $5\,{R_{\rm eff}}$ from \citet{ala16}.

From the $(B-V)$ colours and $K$ magnitudes from NED for the probable members
of the M\,96 subgroup, the relations derived by \citet{bel03} leads to
stellar masses in the range $M_{\star}=(1.5-6) \times 10^{10}\,{\rm M_{\odot}}$,
with $M_{\star}= 2 \times 10^{10}\,{\rm M_{\odot}}$ for NGC\,3377. From the SHMR
derived by \citet{gir20}, the cumulative virial mass is about $(5-8) \times$
$10^{12}\,{\rm M_{\odot}}$, in agreement with the estimation from \citet{kar04b}.
From this expression, the virial mass of NGC\,3377 is $(6\pm3)\times10^{11}$
${\rm M_{\odot}}$.

\citet{hud14} analyse a compilation of galaxies with available information
about the size of their GCS. From a SHMR derived from weak lensing analysis,
they estimate halo masses and define a mean ratio between the mass enclosed
in GCs and the halo mass, $\eta_{\rm M}={M_{\rm GCS}/M_{\rm vir}}$. This parameter
seems to be independent on both the stellar and halo masses of the galaxy,
and it results $\eta_{\rm M}\approx 4\times10^{-5}$ with a large dispersion.
From a different approach, \citet{for16} arrive to $\eta_{\rm M}\approx 3.4$
$\times10^{-5}$, which is also in agreement with the results from \citet{har17},
${\rm log}\eta_{\rm M} =-4.54\pm0.28$, for a sample that spans from ultra
diffuse galaxies to galaxy clusters. Considering for NGC\,3377 the absolute
magnitude $M_V=-19.84$ and Equation\,2 from \citet{har17}, the mean mass of
its GCs is $1.8_{-0.6}^{+1}\times10^5\,{\rm M_{\odot}}$. Hence, the size of the
GCS derived in Section\,\ref{photGCs} implies that $M_{\rm GCS}=(5.9\pm2)$
$\times10^7\,{\rm M_{\odot}}$. Finally, from the ratio $\eta_{\rm M}$ derived
by \citet{har17}, it results $M_{\rm vir}=(2\pm1.5)\times10^{12}\,{\rm M_{\odot}}$.
This estimation is larger than those derived in this paper, and is comparable
with the virial mass calculated for the entire M\,96 subgroup by \citet{kar04b}.
However, the scatter in the parameters propagate to the virial mass estimation,
leading to large uncertainties. Moreover, the ratio $\eta_{\rm M}$ might present
differences between central and satellite galaxies, considering the substantial
mass loss for subhaloes reported in numerical studies
\citep[e.g.][]{gan10,ram16,cor18}.

Hence, the comparison with results and scaling relations from the literature 
are in agreement with the results from this paper, assuming NGC\,3377 as a satellite 
galaxy in the M\,96 group. From this scenario, the isotropic case and that with 
different anisotropies (i.e. $\beta_{\rm GCs}=-0.5$, $\beta_{\rm stel}=0$ and 
$\beta_{\rm PNe}=0.5$) are both plausible and lead to similar results.

\section{Summary}
The aim of this paper was to provide an alternative method to measure
virial masses and the most likely parameters of the mass profile for
intermediate mass galaxies, that usually present reduced populations
of kinematical tracers. Besides the standard use of mass estimators,
an accurate analysis of the mass profiles in these galaxies might
provide clues about the mass accretion processes that lead to the
current mass distribution function in the nearby Universe. This is
particularly relevant for satellite galaxies in dense environments,
that have experienced a considerable mass loss from the moment of
infall. The implementation of Bayesian statistics based on {\it prior}
distributions built from dark matter simulations have proven to be
useful to constraint the parameters of the mass profile. Comparing with
typical spherical Jeans analysis for kinematical tracers, the second
change is the use of the Gauss-Hermite series of the velocity
distribution in the line-of-sight. This was motivated in the reduced
size of the samples and their disperse spatial distribution, that
prevented from grouping them in ranges of galactocentric distance to
estimate velocity dispersions in the line-of-sight.

Observations of NGC\,3377 in two broad bands were downloaded from the
HST/ACS archive, and their photometry was performed to measure GC
candidates and obtain the radial profile of its GCS. Besides, Gemini/GMOS 
long-slit spectroscopic observations for two orientations aligned with 
the major and minor axes of NGC\,3377 were reduced. The velocities and 
dispersions in the line-of-sight of the diffuse stellar population were 
measured at different galactocentric distances, through a dedicated 
algorithm. These observational data set was supplemented with kinematical 
data available from the literature for the diffuse stellar population 
and halo tracers, like globular clusters and planetary nebulae. Several 
options of constant anisotropy were considered. Both NFW and Einasto 
profiles provide consistent results for the mass profile of NGC\,3377. 
From a comprehensive analysis of the observational evidence, the solution 
involving satellite galaxies was preferred, resulting in a virial mass of 
$(3.6\pm0.6) \times 10^{11}\,{\rm M_{\odot}}$. Finally, the comparison 
with independent measurements and scaling relations from the literature 
shows a good agreement.

\section*{Acknowledgments}
The author is grateful with the anonymous referee, whose comments improve the 
article. This work was funded with grants from Consejo Nacional de Investigaciones
Cient\'{\i}ficas y T\'ecnicas de la Rep\'ublica Argentina, Agencia Nacional
de Promoci\'on Cient\'{\i}fica y Tecnol\'ogica, and Universidad Nacional
de La Plata (Argentina). JPC is grateful to Francisco Azpilicueta for useful
comments on statistical issues, and to Ricardo Salinas. Based on observations
obtained at the international Gemini Observatory, a program of NSF's NOIRLab,
which is managed by the Association of Universities for Research in Astronomy
(AURA) under a cooperative agreement with the National Science Foundation on
behalf of the Gemini Observatory partnership: the National Science Foundation
(United States), National Research Council (Canada), Agencia Nacional de
Investigaci\'on y Desarrollo (Chile), Ministerio de Ciencia, Tecnolog\'ia e
Innovaci\'on (Argentina), Minist\'erio da Ci\^{e}ncia, Tecnologia,
Inova\c{c}\~{o}es e Comunica\c{c}\~{o}es (Brazil), and Korea Astronomy and
Space Science Institute (Republic of Korea). Based on observations made with
the NASA/ESA Hubble Space Telescope, obtained from the data archive at the
Space Telescope Science Institute. STScI is operated by the Association of
Universities for Research in Astronomy,  Inc. under NASA contract NAS 5-26555.
This research has made use of the NASA/IPAC Extragalactic Database (NED)
which is operated by the Jet Propulsion Laboratory, California Institute
of Technology, under contract with the National Aeronautics and Space
Administration.

\section*{Data Availability}
The kinematical data reduced in this paper is presented in
Table\,\ref{tab.kinem} at the Appendix. Data from other sources
are available in the corresponding electronic versions of the
papers, \citet{coc09} for the kinematics of the diffuse stellar
population and the planetary nebulae, and \citet{pot13} for the
GCs from the SLUGGS survey. The MDPL2 simulation is publicly
available at the official database of the Multidark project
(\url{https://www.cosmosim.org/}). The reduced images from
ACS/HST for NGC\,3377 can be downloaded from the MAST archive.

\bibliographystyle{mnras}
\bibliography{biblio}

\appendix

\begin{onecolumn}
\section{Definition of the kernel}

In Section\,\ref{sec.jeans} it was stated that the expression for $\overline{V_{\rm LOS}^4}$
can be easily reduced to a double integral in the cases of constant anisotropy
analysed in this paper, assuming a kernel $K^4(R,s)$ which comes from:

\begin{equation}
  K^4(R,s) = s^{2\beta-1} \int_R^s\left[1-2\beta\frac{R^2}{r^2}+\frac{\beta(1+\beta)}{2}\frac{R^4}{r^4}\right]\frac{3\,r^{-2\beta+1}}{\sqrt{r^2-R^2}} dr
\end{equation}

\medskip
\noindent In the isotropic case ($\beta=0$), it results

\begin{equation}
    K_{iso}^4(R,s) = 3\,\sqrt{1-\frac{R^2}{s^2}}
\end{equation}

\medskip
\noindent For the specific values, $\beta=1/2,-1/2$, the expressions are

\begin{equation}
    K_{+1/2}^4(R,s) = \frac{3}{2}\bigg[2\,{\rm tanh^{-1}}\sqrt{1-\frac{R^2}{s^2}}+
    \sqrt{1-\frac{R^2}{s^2}}\left(\frac{1}{4}\frac{R^2}{s^2}-\frac{3}{2}\right) \bigg] \\
\end{equation}
\begin{equation}
    K_{-1/2}^4(R,s) = \frac{3}{2}\bigg[3\frac{R^2}{s^2}\,{\rm tanh^{-1}}\sqrt{1-\frac{R^2}{s^2}}+
    \sqrt{1-\frac{R^2}{s^2}}\left(1-\frac{1}{4}\frac{R^2}{s^2}\right) \bigg] \\
\end{equation}

\medskip
\noindent Finally, for the stationary case with intermediate values of anisotropy,
$-1/2 < \beta < 1/2$ and $\beta\not=0$, 

\begin{eqnarray*}
\begin{aligned}
    K_{est}^4(R,s) =& \frac{3}{2}\left(\frac{s}{R}\right)^{2\beta-1}\bigg\{ \left(\frac{R}{s}\right)^{2\beta-1}\sqrt{1-\frac{R^2}{s^2}}\bigg[\frac{\beta}{2}\frac{R^2}{s^2}-\frac{1}{(\beta-0.5)}\bigg] + \\
    & \bigg[1-I\left(\frac{R^2}{s^2};\beta+0.5,0.5\right)\bigg]B(\beta-0.5,0.5)
    \left(\frac{\beta^2}{2}-2\beta+\frac{15}{8}\right) \bigg\}
\end{aligned}
\end{eqnarray*}

\noindent where $B(a,b) = \int_0^1 t^{a-1}\,(1-t)^{b-1}\,dt$ is the beta function,
$B_{sup}(x;a,b) = \int_x^1 t^{a-1}\,(1-t)^{b-1}\,dt$ represents the incomplete beta
function, and $I(x;a,b)$ is the regularized incomplete beta function, derived
from the previous ones by the relation $I(x;a,b)= B_{sup}(x;a,b)/B(a,b)$.
\end{onecolumn}

\section{Kinematics of NGC\,3377}

\begin{table*}
\begin{center}
  \caption{Kinematical data from the analysis of the GMOS long-slit data,
    measured through p{\sc pxf}.}   
\label{tab.kinem}
\begin{tabular}{cccc|cccc}
  \hline
  \hline
  \multicolumn{4}{c|}{{\rm Major axis}} &\multicolumn{4}{|c}{{\rm Minor axis}}\\
  \hline
\multicolumn{1}{c}{$R_{\rm gal}$} &\multicolumn{1}{c}{$\Delta R_{\rm gal}$} &\multicolumn{1}{c}{$V_{\rm LOS}$} &\multicolumn{1}{c|}{$\sigma_{\rm LOS}$}&\multicolumn{1}{|c}{$R_{\rm gal}$} &\multicolumn{1}{c}{$\Delta R_{\rm gal}$} &\multicolumn{1}{c}{$V_{\rm LOS}$} &\multicolumn{1}{c}{$\sigma_{\rm LOS}$}\\   
\multicolumn{1}{c}{{\rm arcsec}}&\multicolumn{1}{c}{{\rm arcsec}}& \multicolumn{1}{c}{${\rm km\,s^{-1}}$} &\multicolumn{1}{c|}{${\rm km\,s^{-1}}$} & \multicolumn{1}{|c}{{\rm arcsec}}&\multicolumn{1}{c}{{\rm arcsec}}& \multicolumn{1}{c}{${\rm km\,s^{-1}}$} &\multicolumn{1}{c}{${\rm km\,s^{-1}}$}\\   
\hline
-117 & 18 & $875\pm13$ & $105\pm27$ & -- & -- & -- & --\\
-101 & 14 & $727\pm16$ & $69\pm23$ & 	-- & -- & -- & --\\
-88 & 12 & $728\pm30$ & $95\pm21$ & -- & -- & -- & --\\
-78 & 8 & $729\pm24$ & $99\pm24$ & -93 & 16 & $689\pm32$ & $106\pm20$\\
-70 & 8 & $715\pm15$ & $94\pm14$ & -79 & 12 & $697\pm37$ & $93\pm27$\\
-63 & 6 & $739\pm19$ & $102\pm13$ & -69 & 8 & $702\pm28$ & $105\pm37$\\
-57 & 6 & $737\pm15$ & $88\pm10$ & -61 & 8 & $728\pm26$ & $127\pm30$\\
-47 & 6 & $760\pm32$ & $56\pm10$ & -47 & 6 & $698\pm13$ & $120\pm28$\\
-42 & 4 & $769\pm18$ & $68\pm13$ & -42 & 4 & $676\pm15$ & $78\pm23$\\
-38 & 4 & $766\pm13$ & $63\pm12$ & -38 & 4 & $692\pm16$ & $83\pm17$\\
-34 & 4 & $771\pm14$ & $58\pm10$ & -34 & 4 & $698\pm19$ & $76\pm14$\\
-30.5 & 3 & $774\pm15$ & $67\pm8$ & -30.5 & 3 & $691\pm27$ & $78\pm21$\\
-27.5 & 3 & $774\pm19$ & $45\pm7$ & -27.5 & 3 & $683\pm9$ & $86\pm25$\\
-24.5 & 3 & $766\pm16.5$ & $38\pm12$ & -24.5 & 3 & $686\pm11$ & $76\pm17$\\
-21.5 & 3 & $760\pm26$ & $48\pm12$ & -21.5 & 3 & $698\pm8$ & $67\pm16$\\
-19 & 2 & $764\pm15$ & $60\pm17$ & -19 & 2 & $684\pm10$ & $70\pm19$\\
-17 & 2 & $762\pm14$ & $68\pm11$ & -17 & 2 & $689\pm13$ & $82\pm10$\\
-15 & 2 & $760\pm12$ & $81\pm8$ & -15 & 2 & $689\pm13$ & $79\pm13$\\
-13 & 2 & $758\pm11$ & $81\pm7$ & -13 & 2 & $687\pm12$ & $84\pm8$\\
-11 & 2 & $762\pm12$ & $80\pm8$ & -11 & 2 & $691\pm12$ & $77\pm11$\\
-9 & 2 & $764\pm10$ & $88\pm7$ & -9 & 2 & $688\pm11$ & $71\pm15$\\
-7 & 2 & $767\pm9$ & $88\pm8$ & -7 & 2 & $690\pm11$ & $81\pm8$\\
-5 & 2 & $770\pm7$ & $102\pm8$ & -5 & 2 & $688\pm10$ & $86\pm7$\\
-3 & 2 & $765\pm7$ & $113\pm8$ & -3 & 2 & $688\pm8$ & $101\pm7$\\
-1 & 2 & $679\pm5$ & $171\pm9$ & -1 & 2 & $674\pm5$ & $172\pm8$\\
1 & 2 & $648\pm5$ & $165\pm9$ & 1 & 2 & $673\pm5$ & $175\pm8$\\
3 & 2 & $583\pm7$ & $111\pm7$ & 3 & 2 & $686\pm8$ & $103\pm7$\\
5 & 2 & $581\pm8$ & $102\pm7$ & 5 & 2 & $686\pm11$ & $82\pm8$\\
7 & 2 & $582\pm10$ & $91\pm7$ & 7 & 2 & $689\pm12$ & $82\pm7$\\
9 & 2 & $588\pm12$ & $76\pm9$ & 9 & 2 & $685\pm14$ & $77\pm7$\\
11 & 2 & $590\pm11$ & $75\pm11$ & 11 & 2 & $687\pm13$ & $78\pm8$\\
13 & 2 & $593\pm13$ & $71\pm11$ & 13 & 2 & $688\pm14$ & $68\pm19$\\
15 & 2 & $593\pm11$ & $67\pm20$ & 15 & 2 & $690\pm8$ & $73\pm25$\\
17 & 2 & $591\pm20$ & $54\pm17$ & 17 & 2 & $688\pm15$ & $84\pm10$\\
19 & 2 & $593\pm12$ & $73\pm11$ & 19 & 2 & $694\pm9$ & $85\pm22$\\
21.5 & 3 & $588\pm17$ & $50\pm15$ & 21.5 & 3 & $685\pm14$ & $68\pm28$\\
24.5 & 3 & $587\pm23$ & $45\pm13$ & 24.5 & 3 & $696\pm7$ & $103\pm21$\\
27.5 & 3 & $585\pm23$ & $43\pm9$ & 27.5 & 3 & $684\pm8$ & $105\pm31$\\
30.5 & 3 & $589\pm10$ & $57\pm9$ & 30.5 & 3 & $681\pm16$ & $83\pm28$\\
34 & 4 & $591\pm16$ & $60\pm9$ & 34 & 4 & $700\pm11$ & $111\pm22$\\
38 & 4 & $592\pm14$ & $64\pm11$ & 38 & 4 & $709\pm13$ & $98\pm21$\\
42 & 4 & $594\pm19$ & $69\pm20$ & 42 & 4 & $694\pm13$ & $85\pm15$\\
47 & 6 & $600\pm15$ & $84\pm9$ & 47 & 6 & $702\pm16$ & $124\pm37$\\
57 & 6 & $606\pm11$ & $120\pm12$ & 61 & 8 & $692\pm27$ & $119\pm42$\\
63 & 6 & $628\pm20$ & $104\pm13$ & 69 & 8 & $686\pm25$ & $110\pm29$\\
70 & 8 & $638\pm19$ & $106\pm21$ & 79 & 12 & $682\pm30$ & $89\pm23$\\
78 & 8 & $646\pm20$ & $96\pm25$ & 93 & 16 & $693\pm28$ & $111\pm25$\\
88 & 12 & $663\pm15$ & $114\pm46$ & -- & -- & -- & --\\
101 & 14 & $649\pm14$ & $87\pm23$ & -- & -- & -- & --\\
117 & 18 & $693\pm15$ & $119\pm30$ & -- & -- & -- & --\\
\hline
\hline
\end{tabular}
\end{center}
\end{table*}

\end{document}